\documentclass[12pt, a4paper]{article}
\pdfoutput=1
\usepackage{graphicx}
\usepackage{amssymb}
\usepackage{amsmath}
\usepackage{bm}
\usepackage[table,dvipsnames]{xcolor} %for \rowcolors
\usepackage{cite}
\usepackage{multirow}
\usepackage{slashed}
\usepackage{epstopdf}            
\usepackage{epsfig}
\usepackage{here}
\usepackage{comment}
\usepackage{booktabs} %for \toprule
\usepackage{colortbl} %for \rowcolor
\usepackage{wrapfig}
\usepackage{ascmac}
\usepackage{fancybox}  
\usepackage{soul} % strike through text
\allowdisplaybreaks

\renewcommand{\thefootnote}{\fnsymbol{footnote}}
\numberwithin{equation}{section} % Eq.(Sec.eq.)
\def\beq#1\eeq{\begin{align}#1\end{align}}
\newcommand{\ov}{\overline}

\newcommand{\Lc}{{\Lambda_c}}

\renewcommand{\arraystretch}{1.3}
\RequirePackage{xspace}
\usepackage{caption}
\definecolor{BlueViolet}{rgb}{0.2, 0.00, 0.7}
\definecolor{Blue}{rgb}{0.15, 0.00, 0.9}
\definecolor{lightblue}{rgb}{0.15, 0.35, 0.95}
\definecolor{kitgreen}{rgb}{0
, 0.58823 %150/255
, 0.50980 %130/255
}
\usepackage[%dvipdfmx,
colorlinks=true, linkcolor=lightblue,citecolor=lightblue,urlcolor=kitgreen]{hyperref} 

\begin{document}
\sloppy %https://tex.stackexchange.com/questions/9107/how-can-i-make-my-text-never-go-over-the-right-margin-by-always-hyphenating-or-b
\begin{titlepage}
\begin{center}
%%%%%%%%%%%%%%%%%%%%%%%%%%%%%%%%%%%%%%%%%%%
\hfill{KEK--TH--2838}\\
\vskip .3in

{\large{\bf Semileptonic sum rules in heavy-to-light charm decays}}\\
\vskip .3in

% bold applies to math too
\makeatletter\g@addto@macro\bfseries{\boldmath}\makeatother

{ 
Motoi Endo$^{\rm a,b}$,
Syuhei Iguro$^{\rm c,d}$, 
Satoshi Mishima$^{\rm e}$, \\
Takeru Uchiyama$^{\rm f}$ and
Ryoutaro Watanabe$^{\rm g}$
}

\vskip .2in
$^{\rm a}${\it KEK Theory Center, IPNS, KEK, Tsukuba 305--0801, Japan}\\\vspace{4pt}
$^{\rm b}${\it Graduate Institute for Advanced Studies, SOKENDAI, Tsukuba, Ibaraki 305--0801, Japan} \\\vspace{4pt}
$^{\rm c}${\it Institute for Advanced Research (IAR), Nagoya University,\\ Nagoya 464--8601, Japan}\\\vspace{4pt}
$^{\rm d}${\it Kobayashi-Maskawa Institute (KMI) for the Origin of Particles and the Universe, Nagoya University, Nagoya 464--8602, Japan}\\\vspace{4pt}
$^{\rm e}${\it Department of Liberal Arts, Saitama Medical University, Moroyama, Saitama 350-0495, Japan}\\\vspace{4pt}
$^{\rm f}${\it Department of Physics, Nagoya University, Nagoya 464-8602, Japan}\\\vspace{4pt}
$^{\rm g}${\it Institute of Particle Physics and Key Laboratory of Quark and Lepton Physics (MOE), Central China Normal University, Wuhan, Hubei 430079, China}
\end{center}

%%%%%%%%%%%%%%%%%%%%%%%%%
\begin{abstract}
We investigate semileptonic sum rules in heavy-to-light charm decays, motivated by analogous relations in $b \to c$ and $b \to u$ transitions.
Focusing on the $c \to d\overline{\ell}\nu$ decays, $D \to \pi\overline{\ell}\nu$, $D \to \rho\overline{\ell}\nu$, and $\Lambda_c \to n\overline{\ell}\nu$, we examine a relation among their lepton-flavor universality ratios $R_H^{\mu e}$.
Although the charm sum rule is less precise than the relations in bottom-hadron decays, current low-energy and high-$p_T$ constraints on new physics restrict the actual deviation from the relation to below the percent level.
The relation can therefore provide a useful consistency check of charm semileptonic measurements.
As an application, we derive a prediction for the yet-unmeasured ratio $R_n^{\mu e}$ in $\Lambda_c \to n\overline{\ell}\nu$.
%%%%%%%%%%%%%%%%%%%%%%%%%
\end{abstract}
\end{titlepage}
\setcounter{page}{1}
\renewcommand{\thefootnote}{\#\arabic{footnote}}
\setcounter{footnote}{0}
%%%%%%%%%%%%%%%%%%%%%%%%%
% Contents
%%%%%%%%%%%%%%%%%%%%%%%%%
\hrule
\tableofcontents
\vskip .2in
\hrule
\vskip .4in
%%%%%%%%%%%%%%%%%%%%%%%%%

%%%%%%%%%%%%%%%%%%%%%%%%%%%%%%%
\section{Introduction}
\label{sec:intro}
%%%%%%%%%%%%%%%%%%%%%%%%%%%%%%%
Quark and lepton flavor transitions in semileptonic hadron decays provide important tests of the Standard Model (SM) \cite{Glashow:1961tr,Weinberg:1967tq,Salam:1968rm} and sensitive probes of new physics (NP).
While most flavor observables are well described by the SM \cite{PDG2024}, several measurements show deviations from the SM predictions.
These tensions are being scrutinized at ongoing experiments such as LHCb \cite{Cerri:2018ypt}, Belle II \cite{Belle-II:2018jsg}, and BESIII \cite{BESIII:2020nme}.
Future facilities such as FCC-ee \cite{FCC:2025lpp} and STCF \cite{Achasov:2023gey} are also expected to improve the experimental reach.
These efforts motivate both studies of possible NP effects and complementary tests based on relations among different processes.

Such relations have recently been developed in studies of bottom-hadron semileptonic decays \cite{Blanke:2018yud,Blanke:2019qrx,Fedele:2022iib,Duan:2024ayo,Endo:2025fke,Endo:2025cvu,Endo:2025lvy,Endo:2025set,Endo:2026dxr}.
In particular, explicit sum rules have been found among processes such as $B \to D^{(*)}\tau\nu$ and $\Lambda_b\to\Lambda_c\tau\nu$.
A key feature is that certain combinations of the ratios $R_{H_c}={\rm BR}(H_b\to H_c\tau\ov\nu)/{\rm BR}(H_b\to H_c \ell\ov\nu)$ for $\ell=e,\,\mu$ become almost insensitive to NP contributions.
The origin of these relations in $b\to c$ transitions has been clarified in Ref.~\cite{Endo:2025set} from the viewpoint of heavy-quark symmetry.\footnote{
Related sum rules have also been constructed for other $b \to c$ hadron decays, including orbitally excited modes \cite{Iguro:2026} and modes related by flavor SU(3) symmetry \cite{Iguro:2026xgi}.
}

This feature raises the question whether similar relations can be constructed for other quark-flavor transitions.
A numerical study of $b\to u$ semitauonic transitions was performed in Ref.~\cite{Duan:2024ayo}.\footnote{
Such a study is interesting because ${\rm BR}(B\to \pi\tau\ov\nu)$ \cite{Belle:2015qal} and ${\rm BR}(B\to \tau\ov\nu)$ show slight deviations \cite{Crivellin:2025qsq,Belle-II:2025ruy,Abitalk}.
Future Belle~II data are expected to test the relation \cite{Belle-II:2018jsg}.
In addition, heavy-to-light sum rules in $b\to s\nu\ov\nu$ transitions have been discussed in Refs.~\cite{Lee:2025kvf,Kitahara:2026doj}.
}
In contrast to the $b\to c$ case, this heavy-to-light relation is not known to follow from heavy-quark symmetry.
Nevertheless, the numerical study found that the relation remains approximately satisfied.
It is therefore natural to ask whether a similar relation can appear in charm decays.

In this work, we address this question by focusing on $c\to d\ov{\ell}\nu$ transitions.
In particular, we examine whether a predictive sum rule can be constructed for $D \to \pi^- \ov{\ell}\nu$, $D \to \rho^- \ov{\ell}\nu$, and $\Lambda_c\to n \ov{\ell}\nu$.

Experimentally, several branching fractions are already available for the mesonic modes $D \to \pi^- \ov{\ell}\nu$ and $D \to \rho^- \ov{\ell}\nu$.
For the $D \to \pi^-$ channel, branching fractions have been reported for both the electron and muon modes by Belle \cite{Belle:2006idb} and BESIII \cite{BESIII:2015tql,BESIII:2018nzb}, with additional measurements of the electron mode from CLEO \cite{CLEO:2005cuk,CLEO:2007ntr,CLEO:2009svp}.
For the $D \to \rho^-$ channel, the electron mode has been measured by CLEO \cite{CLEO:2005cuk,CLEO:2011ab} and BESIII \cite{BESIII:2018qmf,BESIII:2024lxg}, while the muon mode is currently available only from BESIII \cite{BESIII:2021pvy}.
A direct estimate of the lepton-flavor universality ratio is currently available only for $D \to \pi^-$, where $R^{\mu e}_{\pi}={\rm BR}(D^0\to\pi^-\ov{\mu}\nu)/{\rm BR}(D^0\to\pi^-\ov{e}\nu)$ shows a mild deviation from the SM prediction at the $1.7\,\sigma$ level \cite{BESIII:2018nzb,Riggio:2017zwh}.
In contrast, no experimental result has yet been reported for the baryonic process $\Lambda_c\to n \ov{\ell}\nu$.

On the theoretical side, lattice QCD (LQCD) results are available for $D \to \pi$ \cite{Lubicz:2017syv,Lubicz:2018rfs} and $\Lambda_c\to n$ \cite{Meinel:2017ggx}.
For $D \to \rho$, an early LQCD study treated the $\rho$ meson as a stable particle \cite{Bowler:1994zr}, while a light-cone sum-rule (LCSR) calculation is also available \cite{Wu:2006rd}.
In this work, we use the LCSR input for $D \to \rho$.

With these experimental data and form factor inputs, the sum rule studied here provides a consistency check among the charm semileptonic ratios.
This check will become more useful as precision charm measurements improve \cite{BESIII:2020nme,Achasov:2023gey}, and it directly relates the mesonic ratios to the yet-unmeasured baryonic ratio in $\Lambda_c\to n \ov{\ell}\nu$.

The rest of the paper is organized as follows.
In Sec.~\ref{sec:Pre}, we introduce the operator basis to describe NP contributions and review the decay rates.
In Sec.~\ref{sec:SRviolation}, we describe the sum rule construction and define a cancellation measure.
We also compare the size of the sum rule violation with those in $b\to c l\ov\nu$ and $b\to u l\ov\nu$ decays for $l=e,\mu,\tau$.
In Sec.~\ref{sec:Correction}, we discuss phenomenological implications of the violation under current constraints and present a sum rule prediction for $R_n^{\mu e}$.
Section~\ref{sec:Summary} is devoted to the summary and discussion.

%%%%%%%%%%%%%%%%%%%%%%%%%%%%%%%%%%%%%
\section{Setup}
\label{sec:Pre}
%%%%%%%%%%%%%%%%%%%%%%%%%%%%%%%%%%%%%

%%%%%%%%%%%%%%%%%%%%%%%%%%%%%%%%%%%%%
\subsection{Operator basis}
\label{sec:Heff}
%%%%%%%%%%%%%%%%%%%%%%%%%%%%%%%%%%%%%

We assume that NP contributes to the $c\to d \ov \ell \nu$ transitions. 
The final state contains $\ov\ell\nu$, rather than $\ell\ov\nu$ as in $b\to c l\ov\nu$ decays.
The weak effective Lagrangian is written as 
\begin{align}
 \label{eq:Hamiltonian}
 {\cal {L}}_{\rm{eff}}= -\frac{4  G_FV_{cd}^*}{\sqrt2}\biggl[ (1+C_{V_L}^\ell)O_{V_L}^\ell+C_{S_L}^\ell O_{S_L}^\ell+C_{S_R}^\ell O_{S_R}^\ell+C_{T}^\ell O_{T}^\ell\biggr]\,.
\end{align}
The effective operators are defined as
\begin{align}
 O_{V_L}^\ell &= (\overline{d} \gamma^\mu P_L c)(\overline{\nu}_\ell \gamma_\mu P_L \ell)\,,\,\,\, \,\,
 O_{S_L}^\ell = (\overline{d}  P_L c)(\overline{\nu}_\ell P_{R} \ell)\,,\notag \\
 O_{S_R}^\ell &= (\overline{d}  P_R c)(\overline{\nu}_\ell P_{R} \ell)\,,  \,\,\,\,
 O_{T}^\ell = (\overline{d}  \sigma^{\mu\nu}P_R c)(\overline{\nu}_\ell \sigma_{\mu\nu} P_{R} \ell) \,,
 \label{eq:operator}
\end{align}
where $P_L=(1-\gamma_5)/2$, $P_R=(1+\gamma_5)/2$, and $\sigma^{\mu\nu}=(i/2)[\gamma^\mu,\gamma^\nu]$ with the convention $\sigma^{\mu\nu} \gamma_5 = - (i/2) \epsilon^{\mu\nu\rho\sigma} \sigma_{\rho\sigma}$.
NP effects at high energy scales are encoded in the Wilson coefficients $C_X$.
The normalization factor $4 G_F V_{cd}^*/\sqrt{2}$ corresponds to the SM contribution, which is obtained by setting $C_X=0$ for $X=V_L$, $S_{L,R}$, and $T$.
All light neutrinos are assumed to be left-handed.
Throughout this paper, the Wilson coefficients are evaluated at $\mu=2\,$GeV unless otherwise stated.

%%%%%%%%%%%%%%%%%%%%%%%%%%%%%%%%%%%%%%%%
\subsection{Decay rate}
\label{sec:DR}
%%%%%%%%%%%%%%%%%%%%%%%%%%%%%%%%%%%%%%%%

For $D \to \pi^- \ov \ell \nu$, the differential decay rate is given by
\begin{align}
 \frac{d\Gamma (D \to \pi \ov \ell\nu)}{dq^2} &= 
 \frac{G_F^2 |V_{cd}|^2}{192\pi^3}
 \frac{q^2}{m_{D}^3}\sqrt{\lambda_{\pi}}
 \left( 1 - \rho_\ell^2 \right)^2 
 \notag \\ 
 & \times\bigg\{
 |1 + C_{V_L}^\ell|^2 
 \left[ \left( 1 + \frac{1}{2} \rho_\ell^2 \right) (H_{V,0}^{\pi})^2 + \frac{3}{2} \rho_\ell^2 (H_{V,t}^{\pi})^2 \right] 
 \notag \\ 
 & \quad
 + \frac{3}{2} \, |C_{S_L}^\ell + C_{S_R}^\ell|^2 (H_S^{\pi})^2 
 + 8 \, |C_T^\ell|^2 ( 1 + 2 \rho_\ell^2 ) (H_T^{\pi})^2
 \notag \\ 
 & \quad
 - 3 \, {\rm {Re}}\big[ ( 1 + C_{V_L}^\ell ) (C_{S_L}^\ell + C_{S_R}^\ell )^{*} \big] \rho_\ell H_S^{\pi} H_{V,t}^{\pi} 
 \notag \\ 
 & \quad
 - 12 \, {\rm {Re}}\big[ ( 1 + C_{V_L}^\ell ) C_T^{\ell*} \big] \rho_\ell H_{V,0}^{\pi} H_T^{\pi} 
 \bigg\} \,,
 \label{eq:DDR_D}
\end{align}
where $q^2$ is the invariant mass squared of the lepton system, $\lambda_{H_d}=((m_{H_c}-m_{H_d})^2-q^2)((m_{H_c}+m_{H_d})^2-q^2)$, and $\rho_\ell = m_\ell/\sqrt{q^2}$.
The hadronic amplitudes $H^{\pi}$ are functions of the form factors given explicitly in Appendix~\ref{app:FF}.
Compared with the formulae for $b\to c l\ov\nu$ transitions \cite{Sakaki:2013bfa,Endo:2025fke}, the CP-conjugate $\ov\ell\nu$ final state leads to the opposite sign for the scalar-vector interference terms.
Similarly, the rate for $D \to \rho^- \ov \ell \nu$ is given by
\begin{align}
 & \frac{d\Gamma (D \to \rho \ov \ell\nu)}{dq^2} =
 \frac{G_F^2 |V_{cd}|^2}{192\pi^3}
 \frac{q^2}{m_{D}^3}\sqrt{\lambda_{\rho}}
 \left( 1 - \rho_\ell^2 \right)^2 
 \notag \\ 
 & \qquad
 \times\bigg\{
 |1 + C_{V_L}^\ell|^2 \left[ \left( 1 + \frac{1}{2} \rho_\ell^2 \right) \big[ ( H_{V,0}^{\rho} )^2 + ( H_{V,+}^{\rho} )^2 + ( H_{V,-}^{\rho} )^2 \big] + \frac{3}{2} \rho_\ell^2 ( H_{V,t}^{\rho} )^2 \right] 
 \notag \\ 
 & \qquad\quad
 + \frac{3}{2} \, |C_{S_L}^\ell - C_{S_R}^\ell|^2 ( H_{P}^{\rho} )^2 
 + 8 \, |C_T^\ell|^2 ( 1+ 2 \rho_\ell^2 ) \big[ ( H_{T,0}^{\rho} )^2 + ( H_{T,+}^{\rho} )^2 + ( H_{T,-}^{\rho} )^2 \big] 
 \notag \\ 
 & \qquad\quad
 + 3 \, {\rm {Re}}\big[ (1 + C_{V_L}^\ell ) (C_{S_L}^\ell - C_{S_R}^\ell )^{*} \big] \rho_\ell H_{P}^{\rho} H_{V,t}^{\rho} 
 \notag \\ 
 & \qquad\quad
 - 12 \, {\rm {Re}}\big[ (1 + C_{V_L}^\ell) C_T^{\ell*} \big] \rho_\ell \left( H_{V,0}^{\rho} H_{T,0}^{\rho} + H_{V,+}^{\rho} H_{T,+}^{\rho} - H_{V,-}^{\rho} H_{T,-}^{\rho} \right) 
 \bigg\} \,.
 \label{eq:DDR_Ds}
\end{align}
Analogously, the differential decay rate for $\Lambda_c\to n\ov \ell \nu$ is obtained based on Refs.~\cite{Datta:2017aue,Bernlochner:2018bfn}
\begin{align}
 \label{eq:DDR_Lambda}
 & \frac{d\Gamma(\Lambda_c\to n\ov \ell\nu)}{dq^2} =
\frac{G_F^2 |V_{cd}|^2}{384\pi^3}
 \frac{q^2}{m_{\Lambda_c}^3}\sqrt{\lambda_{n}}
 \left( 1 - \rho_\ell^2 \right)^2 
 \\ 
 & \quad
 \times\bigg\{
 |1 + C_{V_L}^\ell|^2 \bigg[ \bigg( 1 + \frac{1}{2} \rho_\ell^2 \bigg) \big[ (H_{V,0+}^{n})^2 + (H_{V,0-}^{n})^2 + (H_{V,1+}^{n})^2 + (H_{V,1-}^{n})^2  \big]
 \notag \\ 
 & \qquad\qquad\qquad\qquad\quad
 + \frac{3}{2} \rho_\ell^2 \big[ (H_{V,t+}^{n})^2 + (H_{V,t-}^{n})^2 \big] \bigg]
 \notag \\ 
 & \qquad
 + 3 \Big[ |C_{S_L}^\ell+C_{S_R}^\ell|^2 (H_{S}^{n})^2 + |C_{S_L}^\ell-C_{S_R}^\ell|^2 (H_{P}^{n})^2 \Big]
 \notag \\ 
 & \qquad
 + 8 \, |C_T^\ell|^2 (1 + 2\rho_\ell^2) \Big[ (H_{T,0+}^{n})^2 + (H_{T,0-}^{n})^2 + (H_{T,1+}^{n})^2 + (H_{T,1-}^{n})^2 \Big]
 \notag \\ 
 & \qquad
 - 3 \, {\rm {Re}}\big[ (1+C_{V_L}^\ell) (C_{S_L}^\ell+C_{S_R}^\ell)^* \big] \rho_\ell (H_{V,t+}^{n} + H_{V,t-}^{n}) H_{S}^{n} 
 \notag \\ 
 & \qquad
 - 3 \, {\rm {Re}}\big[ (1+C_{V_L}^\ell) (C_{S_L}^\ell-C_{S_R}^\ell)^* \big] \rho_\ell (H_{V,t+}^{n} - H_{V,t-}^{n}) H_{P}^{n}
 \notag \\ 
 & \qquad
 + 12 \, {\rm {Re}}\big[(1+C_{V_L}^\ell) C_T^{\ell*}\big] \rho_\ell
 \big( H_{V,0+}^{n} H_{T,0+}^{n} + H_{V,0-}^{n} H_{T,0-}^{n} + H_{V,1+}^{n} H_{T,1+}^{n} + H_{V,1-}^{n} H_{T,1-}^{n} \big)
 \bigg\} \,.
 \notag
\end{align}
Using these formulae, we construct the charm sum rule and evaluate its cancellation properties in the next section.

%%%%%%%%%%%%%%%%%%%%%%%%%%%%%%%%%%%%%%%%
\section{Charm sum rule and cancellation}
\label{sec:SRviolation}
%%%%%%%%%%%%%%%%%%%%%%%%%%%%%%%%%%%%%%%%

We now construct the sum rule for the $c\to d$ semileptonic ratios and quantify its violation.
Since decays with a $\tau$ lepton in the final state are kinematically forbidden in charm semileptonic decays, we consider lepton-flavor universality ratios between the muon and electron modes.
In general, NP may contribute to the muon mode through $C_X^\mu$ and/or to the electron mode through $C_X^e$.
For the electron mode, however, the interference terms with scalar and tensor operators are strongly suppressed by the charged-lepton mass.
We focus on NP contributions to the muon mode in the following analysis and comment on other possibilities later.

Using the form factor inputs from LQCD for $D\to\pi$ \cite{Lubicz:2017syv,Lubicz:2018rfs} and $\Lambda_c\to n$ \cite{Meinel:2017ggx}, and those from LCSR for $D\to\rho$ \cite{Wu:2006rd}, we integrate the differential decay rates in the last section over the phase space.
We define
\begin{align}
R_{H_d}^{\mu e} =
\frac{{\rm BR}(H_c\to H_d\,\ov{\mu}\nu)}
     {{\rm BR}(H_c\to H_d\,\ov{e}\nu)}\,,
\end{align}
where $H_c$ and $H_d$ denote the charm and down hadrons, respectively.
Since NP contributions are assumed only in the muon mode, the normalized ratios $R_{H_d}^{\mu e}/R_{H_d}^{\mu e,{\rm SM}}$ are equivalent to the normalized muon-mode branching fractions and can be written as
\begin{align}
\frac{R_X^{\mu e}}{R_X^{\mu e,{\rm SM}}}
=
\sum_{ij} a^X_{ij}\mathcal{D}_i^\mu \mathcal{D}_j^{\mu *}\,,
\end{align}
with $X=\pi,\rho,n$ and $i,j=V_L,S_L,S_R,T$.
Here $\mathcal{D}_{V_L}^\mu = 1+C_{V_L}^\mu$ and $\mathcal{D}_i^\mu=C_i^\mu$ for $i=S_L,S_R,T$.
We obtain the numerical formulae
\begin{align}
\label{eq:Rpimuon}
\frac{R_{\pi}^{\mu e}}{R_{\pi}^{{\mu e},\,\rm SM}}& =  |1+C^{\mu}_{V_L}|^2 
-0.39  \,{\rm{Re}}[ (1 +C^\mu_{V_L}) C_{S_R}^{\mu*}]
-0.39  \,{\rm{Re}}[ (1 +C^\mu_{V_L}) C_{S_L}^{\mu*} ]
\nonumber \\[0.5em]
&+3.94  \, {\rm{Re}}[C^\mu_{S_L} C_{S_R}^{\mu*} ]
+1.97  \, (|C^\mu_{S_L}|^2 + |C^\mu_{S_R}|^2)
\nonumber \\[0.5em]
&+0.47 \,{\rm{Re}}[ (1 +C^\mu_{V_L})  C_T^{\mu*}] +1.16  \, |C^\mu_T|^2\,,\\
%%%%%
\frac{R_{\rho}^{\mu e}}{R_{\rho}^{{\mu e},\,\rm SM}}& =  |1+C^\mu_{V_L}|^2 
-0.075  \,{\rm{Re}}[ (1 +C^\mu_{V_L}) C_{S_R}^{\mu*}]
+0.075  \,{\rm{Re}}[ (1 +C^\mu_{V_L}) C_{S_L}^{\mu*} ]
\nonumber \\[0.5em]
&-0.21  \, {\rm{Re}}[C^\mu_{S_L} C_{S_R}^{\mu*} ]
+0.11  \, (|C^\mu_{S_L}|^2 + |C^\mu_{S_R}|^2) 
\nonumber \\[0.5em]
&-2.02 \,{\rm{Re}}[ (1 +C^\mu_{V_L})  C_T^{\mu*}] +15.8  \, |C^\mu_T|^2\,,\\
%%%%%
\label{eq:Rnmuon}
\frac{R_{n}^{\mu e}}{R_{n}^{{\mu e},\,\rm SM}}& =   |1+C^\mu_{V_L}|^2 
-0.16  \,{\rm{Re}}[ (1 +C^\mu_{V_L}) C_{S_R}^{\mu*}]
-0.043  \,{\rm{Re}}[ (1 +C^\mu_{V_L}) C_{S_L}^{\mu*} ]
\nonumber \\[0.5em]
&+0.35  \, {\rm{Re}}[C^\mu_{S_L} C_{S_R}^{\mu*} ]
+0.46  \, (|C^\mu_{S_L}|^2 + |C^\mu_{S_R}|^2) 
\nonumber \\[0.5em]
&-0.38 \,{\rm{Re}}[ (1 +C^\mu_{V_L})  C_T^{\mu*}] +7.82  \, |C^\mu_T|^2\,.
\end{align}
Hereafter, the superscript ``$\mu e$'' is omitted unless needed.
Compared with the formulae for the $b\to c$ and $b\to u$ cases in Appendix~\ref{app:bulnu}, the numerical coefficients in the $c\to d$ case are typically larger by a factor of $\sim 1.5$--$4$.
The overall pattern, however, is similar to that in the bottom decays.
The interference terms between $C_{V_L}$ and the scalar or tensor coefficients are relatively small because they are suppressed by $\rho_\ell=m_\ell/\sqrt{q^2}$.
In contrast, the $|C_T|^2$ terms tend to be large, reflecting the factor of $8$ multiplying the tensor contribution in the differential decay rates in Eqs.~\eqref{eq:DDR_D}--\eqref{eq:DDR_Lambda}.

We construct a linear combination of the three normalized ratios, which can be written as
\begin{align}
  \delta
   =\frac{R_{n}}{R_{n}^{\rm SM}}-\alpha \frac{R_{\pi}}{R_{\pi}^{\rm SM}}-\beta \frac{R_{\rho}}{R_{\rho}^{\rm SM}} 
   =\sum_{kl}\delta_{kl} \mathcal{D}_k^\mu \mathcal{D}_l^{\mu *}\, ,
\label{eq:SRdelta}
\end{align}
where $k,l=V_L,\,S_L,\,S_R,\,T$.
If an exact sum rule exists, one can choose $\alpha$ and $\beta$ independently of the Wilson coefficients such that $\delta$ vanishes for arbitrary values of the coefficients.
Equivalently, all coefficients $\delta_{kl}$ vanish simultaneously.
We impose $\alpha+\beta=1$, so that the vector contribution proportional to $|\mathcal{D}_{V_L}^\mu|^2$ cancels in $\delta$.
The remaining nonzero coefficients $\delta_{kl}$ then quantify the residual violation of the sum rule for each operator combination.

In the sum rule for $b\to c$ semileptonic decays, the coefficients $\alpha$ and $\beta$ are fixed by heavy-quark symmetry in the small-velocity limit \cite{Endo:2025lvy}.
In contrast, no analogous symmetry argument is available for the heavy-to-light transitions considered here.
As a first diagnostic, we use a reference choice in which one operator combination is removed from $\delta$, following a prescription developed in Refs.~\cite{Blanke:2018yud,Blanke:2019qrx,Fedele:2022iib,Duan:2024ayo}.
For a chosen operator combination $mn$ with $m,n=V_L,\,S_L,\,S_R,\,T$, we fix $\alpha$ and $\beta$ by requiring $\delta_{mn}=0$.
This yields
\begin{align}
\label{eq:KITSR}
 \alpha_{mn}=\frac{a_{mn}^{n}-a_{mn}^{\rho}}{a_{mn}^{\pi}-a_{mn}^{\rho}}, \qquad
 \beta_{mn}=\frac{a_{mn}^{\pi}-a_{mn}^{n}}{a_{mn}^{\pi}-a_{mn}^{\rho}}\,.
\end{align}
Several choices of $mn$ are possible.
Hereafter, the label $mn$ specifies the operator combination for which the corresponding coefficient $\delta_{mn}$ is zero.

The coefficients $\delta_{kl}$ (for $kl \neq mn$) measure the residual violation in the sum rule.
To quantify the efficiency of the cancellation among the three terms in Eq.~\eqref{eq:SRdelta}, we introduce the cancellation measure \cite{Iguro:2026}
\begin{align}
    \epsilon_{kl} = 
    \frac{|\delta_{kl}|}
    { {\rm{max}}\big[\,|a_{kl}^{n}|,\,|\alpha\,a_{kl}^{\pi}|,\,|\beta\,a_{kl}^{\rho}|\, \big] }\,\,.
\end{align}
In the absence of a special relation among the three decay ratios, one expects $\epsilon_{kl}$ to be of order unity.
A value $\epsilon_{kl}\ll1$ therefore indicates an efficient cancellation.
If this occurs for all operator combinations, the sum rule is well satisfied.

%%%%%%%%%%%%%%%%%%%%%%%%%%%%%%%%%%%%%%%%
\begin{table}[t]
  \renewcommand{\arraystretch}{1.25}
  \centering
  \begin{tabular}{ccccccc} 
    & $\delta_{V_L S_R}$& $\delta_{SS}$ 
    & $\delta_{S_R S_L}$  & $\delta_{V_LT}$ & $\delta_{TT}$& $\alpha$  
    \\ \hline
%%%%%%%
    $c \to d\bar\mu\nu$  
    & $3.2\times 10^{-4}$ & $-0.11$  & $-0.48$ & $1.01$ & $-4.26$  & $0.25$
  \\\hline
%%%%%%%
    $b \to c \mu\bar\nu$ 
    & $4.7\times10^{-3}$  & $-1.2\times10^{-3}$ & $-5.9\times10^{-3}$  & $0.028$ & $0.35$ & $0.27$
    \\
%%%%%%%
    $b \to c \tau\bar\nu$ 
    & $-0.021$  & $-6.3\times 10^{-3}$ & $-0.012$  & $0.13$ & $0.23$ & $0.27$ 
  \\\hline
%%%%%%%
    $b \to u\mu\bar\nu$
    & $-7.9\times10^{-3}$  & $-0.14$ & $-0.33$  & $0.030$ & $-0.56$  & $0.32$
    \\
%%%%%%%
    $b \to u\tau\bar\nu$  
    &  $-0.067$ &  $-0.10$ &  $-0.20$ &  $0.20$ &  $-0.77$ & $0.30$ 
\\\hline
  \end{tabular}
  \caption{
  Sum rule violation coefficients $\delta_{kl}$ for various transitions, evaluated with $\alpha=\alpha_{V_L S_L}$ such that $\delta_{V_L S_L}=0$. 
  The last column shows the corresponding value of $\alpha_{V_L S_L}$.
  }
  \label{tab:delta}
\end{table}
%%%%%%%%%%%%%%%%%%%%%%%%%%%%%%%%%%%%
%%%%%%%%%%%%%%%%%%%%%%%%%%%%%%%%%%%%
\begin{table}[t]
  \centering
  \renewcommand{\arraystretch}{1.3}
  \begin{tabular}{cccccc}     
  \noalign{\vskip -0.4ex}
  \noalign{\vskip -0.7ex}
    & $\epsilon_{V_L S_R}$
    & $\epsilon_{SS}$ 
    & $\epsilon_{S_R S_L}$  & $\epsilon_{V_LT}$ & $\epsilon_{TT}$  
    \\ \hline
%%%%%%%
    $c \to d\bar\mu\nu$  
    &  $2.1\times 10^{-3}$
    &  $0.23$
    &  $0.49$
    & $0.67$ & $0.36$  
    \\\hline
%%%%%%%
    $b \to c \mu\bar\nu$ 
    & $0.014$ 
    & $0.031$
    & $0.12$
    & $0.082$ & $0.030$ 
    \\
%%%%%%%
    $b \to c \tau\bar\nu$ 
    & $0.045$ 
    & $0.022$
    & $0.022$
    & $0.033$ & $0.019$ 
    \\\hline
%%%%%%%
    $b \to u\mu\bar\nu$ 
    & $0.16$ 
    & $0.24$
    & $0.31$
    & $0.24$ & $0.06$ 
    \\ 
%%%%%%%
    $b \to u\tau\bar\nu$ 
    &  $0.12$
    &  $0.19$
    &  $0.22$
    &  $0.12$  & $0.08$ 
\\ \hline
  \end{tabular}
  \caption{
  Cancellation measures $\epsilon_{kl}$ for various transitions, evaluated with the same choice $\alpha=\alpha_{V_L S_L}$ as in Table~\ref{tab:delta}.
 }
  \label{tab:epsilon}
\end{table}
%%%%%%%%%%%%%

\begin{figure*}[t]
\begin{center}
\includegraphics[width=0.43\textwidth]{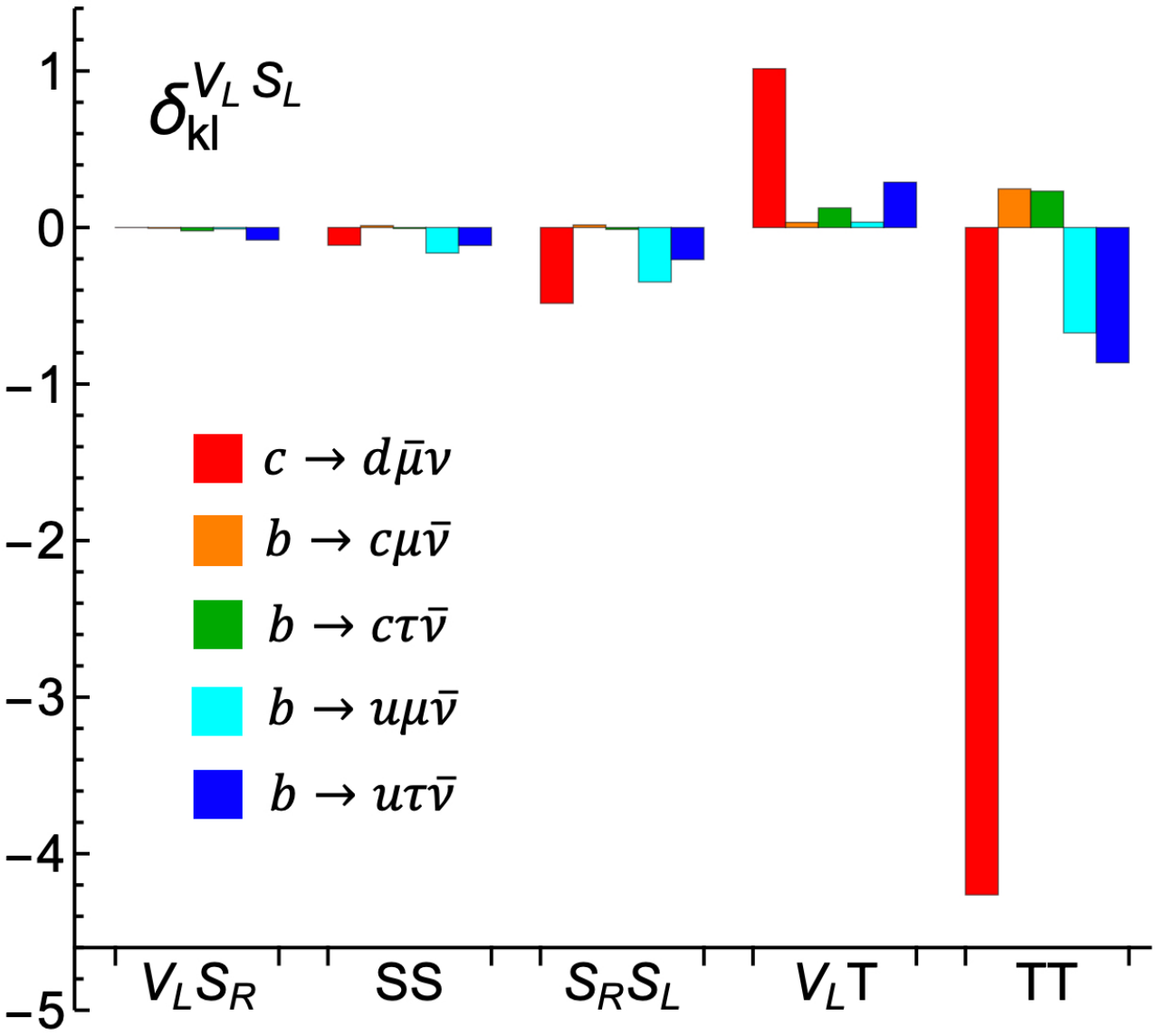}~~~~
\includegraphics[width=0.43\textwidth]{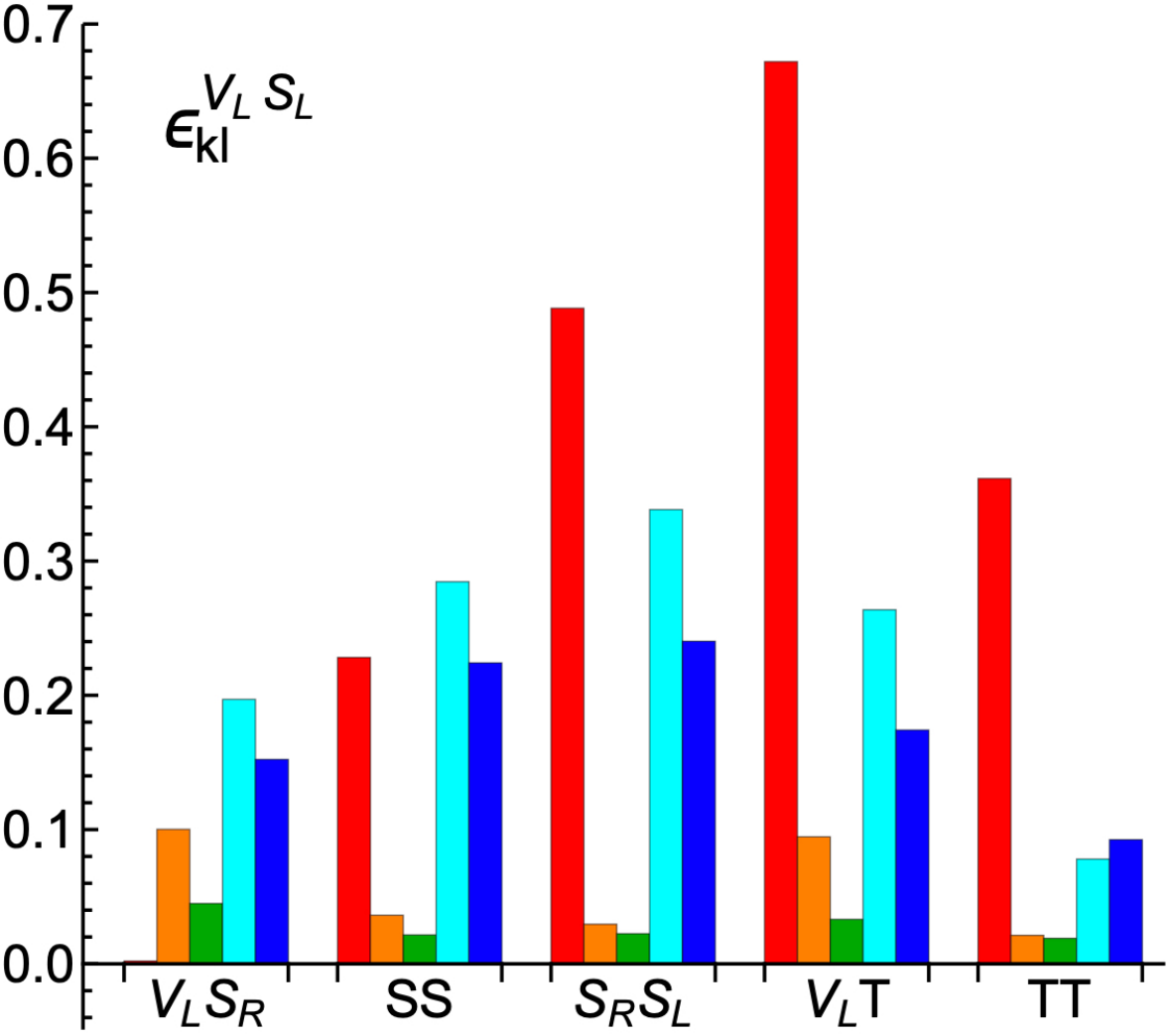}
\caption{
   \label{fig:deleps}
Sum rule violation coefficients $\delta_{kl}$ (left) and cancellation measures $\epsilon_{kl}$ (right) for $c \to d\bar{\mu}\nu$ (red), $b \to c\mu\bar{\nu}$ (orange), $b \to c\tau\bar{\nu}$ (green), $b \to u\mu\bar{\nu}$ (cyan), and $b \to u\tau\bar{\nu}$ (blue). 
The superscript of $\delta$, $V_LS_L$, indicates that the sum rule coefficient $\alpha$ is determined such that the $V_LS_L$ term is eliminated.
} 
\end{center}
\vspace{-3mm}
\end{figure*}

We evaluate $\delta_{kl}$ and $\epsilon_{kl}$ for the reference choice of $\alpha$.
The results for the $c\to d$ transition are shown in the first row of Tables~\ref{tab:delta} and \ref{tab:epsilon}.
They are compared with the results for the $b\to c$ transitions in the second and third rows and those for the $b\to u$ transitions in the fourth and fifth rows.
The comparison is visualized in Fig.~\ref{fig:deleps}.\footnote{
The normalization in $\epsilon_{kl}$ removes the overall size of each operator contribution.
It is therefore better suited than $\delta_{kl}$ for comparing the efficiency of the cancellation among different transitions.
This is useful in particular for the interference terms, whose coefficients are affected by the lepton-mass suppression through $\rho_\ell=m_\ell/\sqrt{q^2}$.
}
Following Ref.~\cite{Duan:2024ayo}, we take $mn=V_LS_L$.
With this convention, $\delta_{V_LS_L}=0$ by construction.
For all transitions, the coefficient $\alpha_{V_LS_L}$, denoted by $\alpha$ in Table~\ref{tab:delta}, clusters around $0.3$.
Table~\ref{tab:epsilon} shows that the $c\to d$ sum rule still exhibits nontrivial cancellations, with $\epsilon_{kl}<1$ for all operator combinations.
In the $c\to d$ case, however, several residual violations are larger than those in the $b\to c$ and $b\to u$ transitions, as shown in Tables~\ref{tab:delta} and \ref{tab:epsilon}.
This is particularly visible for $kl=S_RS_L$, $V_LT$, and $TT$.
Although $\delta_{TT}$ is large, the corresponding value of $\epsilon_{TT}$ remains below unity, indicating that a cancellation is still present in the tensor contribution.
The large residual $\delta_{TT}$ mainly reflects the large tensor coefficients $a_{TT}^X$ in Eqs.~\eqref{eq:Rpimuon}--\eqref{eq:Rnmuon}.

%%%%%%%%%%%%%%%%%%%%%%%%%%%%%%%%%%%%%%%%%%%%%%%%%
%%%%%%%%%%%%%%%%%%%%%%%%%%%%%%%%%%%%%%%%%%%%%%%%%
%%%%%%%%%%%%%%%%%%%%%%%%%%%%%%%%%%%%%%%%%%%%%%%%%
\begin{figure*}[t]
\begin{center}
\includegraphics[width=0.47\textwidth]{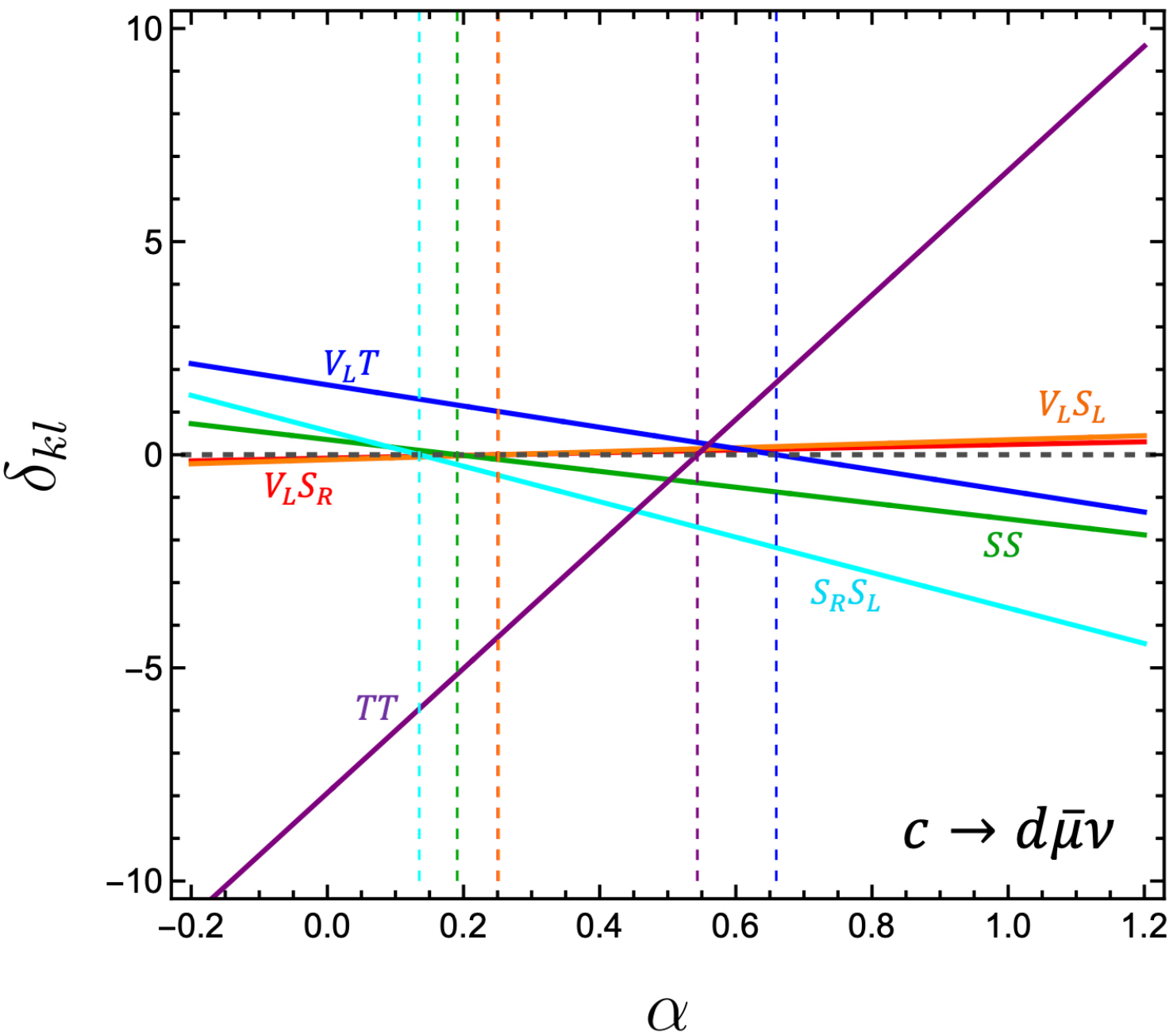}~~~
\includegraphics[width=0.46\textwidth]{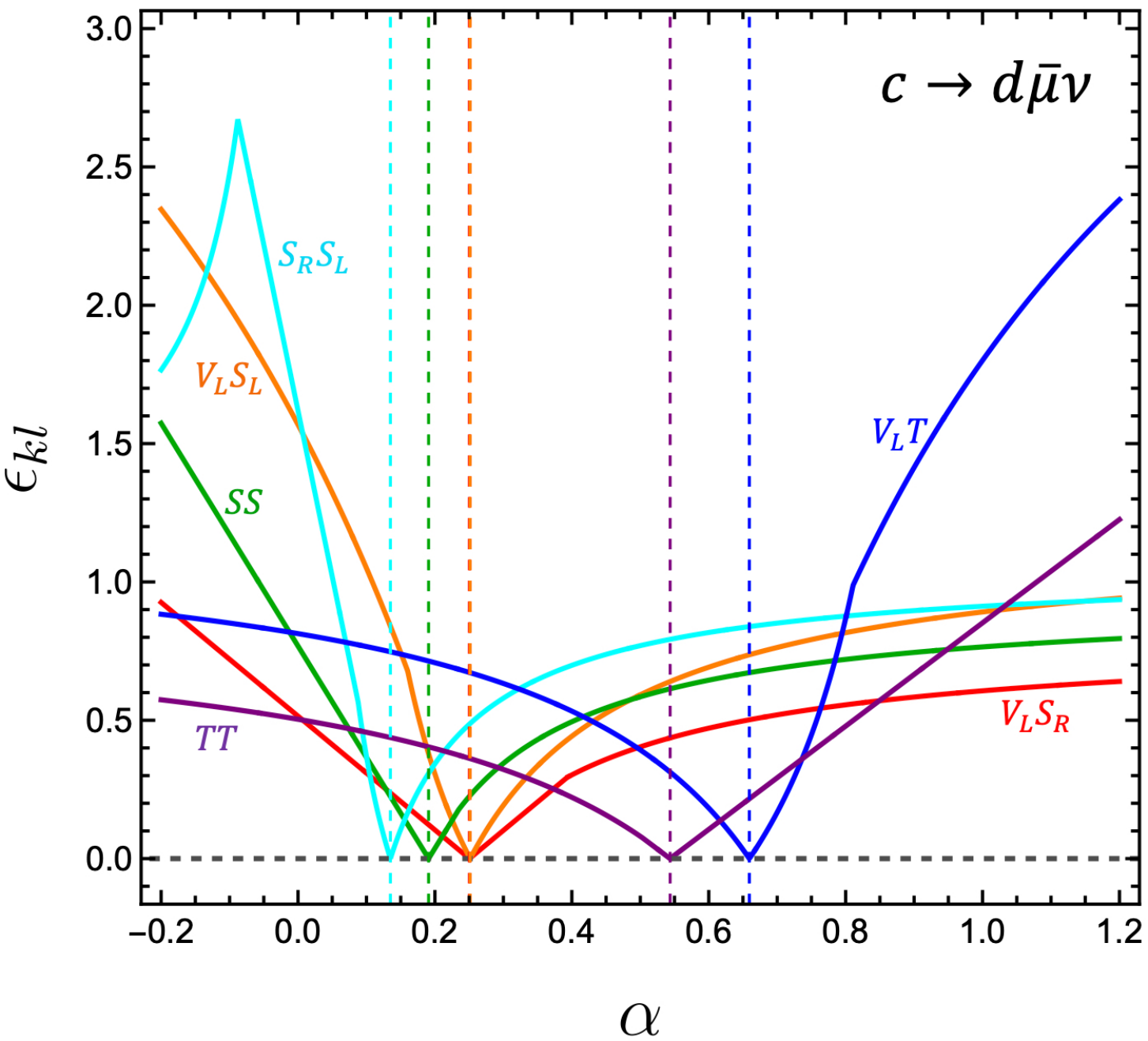}
\caption{
   \label{fig:deleps_alpha_cd}
   The $\alpha$ dependence of $\delta_{kl}$ (left) and $\epsilon_{kl}$ (right) for $c \to d\bar{\mu}\nu$ with $\beta=1-\alpha$. 
   The vertical dashed lines show the values $\alpha_{mn}$ obtained from Eq.~\eqref{eq:KITSR}, where the corresponding $\delta_{mn}$ vanishes.
} 
\end{center}
\vspace{-3mm}
\end{figure*}
%%%%%%%%%%%%%%%%%%%%%%%%%%%%%%%%%%%%%%%%%%%%%%%%%

The choice of $\alpha$ is not unique.
To assess how this choice affects the cancellation pattern, we now treat $\alpha$ as a free parameter while keeping $\alpha+\beta=1$.
This condition continues to remove the common $|\mathcal{D}_{V_L}^\mu|^2$ contribution, and the residual coefficients $\delta_{kl}$ are then obtained from Eq.~\eqref{eq:SRdelta}.
Figure~\ref{fig:deleps_alpha_cd} shows the $\alpha$ dependence of $\delta_{kl}$ and $\epsilon_{kl}$ for $c\to d\ov\mu\nu$.\footnote{
The analogous plots for $b\to c l\ov\nu$ and $b\to u l\ov\nu$ are given in Appendix~\ref{app:al_vari}.
}
The dashed vertical lines indicate the values $\alpha_{mn}$ obtained from Eq.~\eqref{eq:KITSR}, where the corresponding $\delta_{mn}$ vanishes.
In the left panel, $\delta_{TT}$ becomes large in the large $\alpha$ region, while the other residuals remain comparatively small over the range shown.
The right panel shows that the scalar and tensor contributions prefer different regions, with efficient scalar cancellations around $\alpha \approx 0.2$ and tensor cancellations around $\alpha \approx 0.5$.\footnote{
The curves are not always smooth because the dominant term in the denominator changes.
}
All displayed $\epsilon_{kl}$ are below unity in the range $0.1\lesssim\alpha\lesssim0.8$, but no single value of $\alpha$ realizes equally strong cancellations for all operator combinations.
This behavior contrasts with the bottom transitions, where the preferred values of $\alpha$ are more closely aligned, as shown in Appendix~\ref{app:al_vari}.

In summary, the analysis in this section shows that a cancellation in the sum rule construction is present also in the $c\to d$ case, although it is weaker than that in the bottom transitions.
This conclusion is supported both by the reference choice $\delta_{V_LS_L}=0$ and by the scan over $\alpha$.
In particular, the values of $\alpha$ preferred by the scalar and tensor contributions are more widely separated in the $c\to d$ case.
As a result, a single value of $\alpha$ cannot suppress all residual violations as efficiently as in the $b\to c$ and $b\to u$ cases.
Further work is needed to determine whether this pattern reflects a genuine theoretical structure or an accidental numerical feature.
In addition, we have not included form factor uncertainties in this analysis.
This limitation is particularly relevant for $D\to\rho$, where improved lattice determinations would be useful for a more quantitative assessment.

%%%%%%%%%%%%%%%%%%%%%%%%%%%%%%%%%%%%%%%%%%%
\section{Phenomenological implications}
\label{sec:Correction}
%%%%%%%%%%%%%%%%%%%%%%%%%%%%%%%%%%%%%%%%%%%

We discuss the phenomenological implications of the charm sum rule.
Although the cancellation at the coefficient level in the previous section is weaker than that in the bottom transitions, the actual violation of the sum rule also depends on the allowed size of the Wilson coefficients.
We therefore combine the $\alpha$ scan described above with the current constraints on the Wilson coefficients.
As in the previous section, we assume NP contributions only in the muon mode.

%%%%%%%%%%%%%%%%%%%%%%%%%%%%%%%%%%%%%%%%%%%
\subsection{Experimental constraint}
\label{sec:WCconstraint}
%%%%%%%%%%%%%%%%%%%%%%%%%%%%%%%%%%%%%%%%%%%

To estimate the phenomenological size of the residual sum rule violation, we first summarize the current constraints on the Wilson coefficients relevant to $c\to d\ov\mu\nu$ transitions.

The leptonic decay $D^-\to\mu\ov\nu$ gives a stringent low-energy constraint especially on the scalar contributions.
This sensitivity comes from the helicity enhancement of the pseudoscalar matrix element relative to the SM vector-current contribution.
Using the experimental value of ${\rm BR}(D^-\to \mu\ov\nu)$ \cite{PDG2024}, together with $|V_{cd}|\simeq0.225(4)$ and $f_D=0.212(1)\,$GeV \cite{HeavyFlavorAveragingGroupHFLAV:2024ctg, Carrasco:2014poa, Bazavov:2017lyh, FlavourLatticeAveragingGroupFLAG:2024oxs}, we obtain the $2\,\sigma$ allowed range
\begin{align}
   -0.0035\le C_{S_R}- C_{S_L}\le0.00081\,.
\end{align}
The same observable gives a weaker constraint on $C_{V_L}$ than the high-$p_T$ bound discussed below, and we do not use it in the following estimates.

Complementary constraints are obtained from collider experiments.
At the LHC, events with a high-$p_T$ lepton impose stringent bounds on the NP contributing to semileptonic decays.
Reference \cite{Fuentes-Martin:2020lea} reinterpreted ATLAS and CMS results with an integrated luminosity  up to 139\,fb$^{-1}$ \cite{ATLAS:2019lsy,CMS:2018hff} and derived the following bounds on the Wilson coefficients at $\mu=2\,$GeV including renormalization group effects,
\begin{align}
    -0.0085\le C_{V_L}\le 0.012\,,\qquad
    |C_{S_{L,R}}|\le0.023\,,\qquad
    |C_T|\le 0.0052\,.
   \label{eq:LHC_constraint}
\end{align}
These bounds apply to single-operator NP scenarios, in which a single operator describing NP contributions in Eq.~\eqref{eq:Hamiltonian} is assumed to be nonzero while all others vanish.
For scalar operators, the high-$p_T$ bounds are weaker than the $D^-\to\mu\ov\nu$ constraint on the combination $C_{S_R}-C_{S_L}$.
However, the direction $C_{S_R}=C_{S_L}\equiv C_S$ is not constrained by the decay.
For this aligned direction, the high-$p_T$ constraint gives
\begin{align}
|C_S|\le0.016\,.
\end{align}
This follows from the fact that interference effects among different NP operators are negligible in the high-$p_T$ region.

Within the allowed ranges of the Wilson coefficients, $R^{\mu e}_X/R^{\mu e,\,{\rm SM}}_X$ can shift by about $\pm2\%$ when only $C_{V_L}$ is varied, and by about $\pm1\%$ when one of the scalar or tensor coefficients is varied.
The residual violation $\delta$ in the sum rule is reduced compared with these individual shifts because NP contributions cancel among the three decay modes.
In particular, the contribution proportional to $|\mathcal{D}_{V_L}^\mu|^2$ vanishes once $\alpha+\beta=1$ is imposed.
In the following analysis, we take these constraints into account when estimating the possible size of the sum rule violation.

%%%%%%%%%%%%%%%%%%%%%%%%%%%%%%%%%%%%%%%%%%%
\subsection{Phenomenology}
\label{sec:predict}
%%%%%%%%%%%%%%%%%%%%%%%%%%%%%%%%%%%%%%%%%%%

%%%%%%%%%%%%%%%%%%%%%%%%%%%%%%%%%%%%%%%%%%%%%%%%%
\begin{figure*}[t]
\begin{center}
\includegraphics[width=0.6\textwidth]{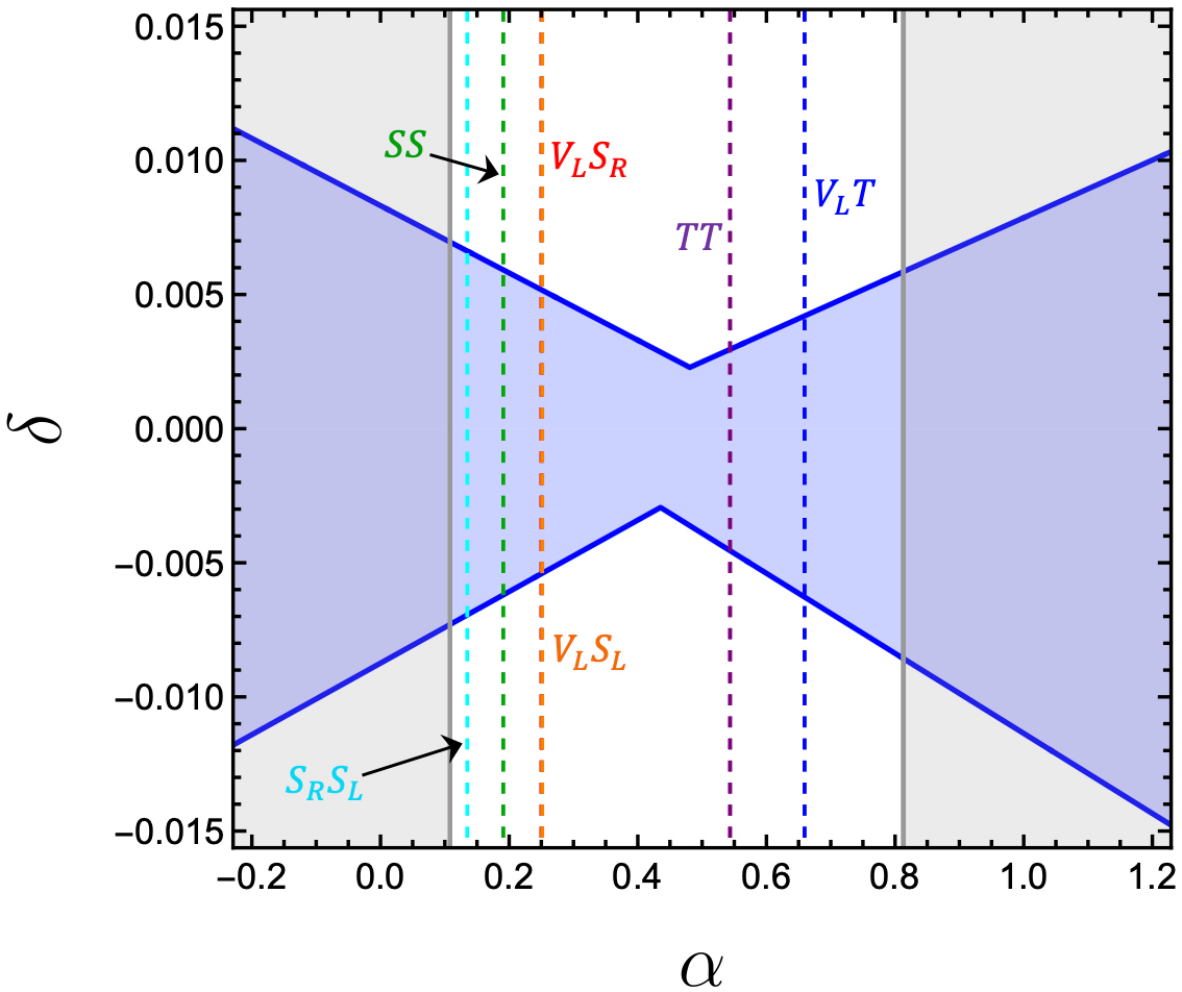}
\caption{
   \label{fig:Deldel}
   Allowed range of the residual sum rule violation $\delta$ as a function of $\alpha$, taking into account the constraints on $C_X$.
   The vertical dashed lines show the values $\alpha_{mn}$ obtained from Eq.~\eqref{eq:KITSR}.
   The gray region corresponds to the range where at least one of the cancellation measures satisfies $\epsilon_{kl}>1$, indicating that the cancellation is not efficient for all operator combinations.
} 
\end{center}
\vspace{-3mm}
\end{figure*}
%%%%%%%%%%%%%%%%%%%%%%%%%%%%%%%%%%%%%%%%%%%%%%%%%

We investigate the actual size of violations in the sum rule.
Even though several $\delta_{kl}$ are numerically large, the physical impact of the violation $\delta$ depends on the size of the accompanying Wilson coefficients.
If $\delta$ is small, the sum rule becomes insensitive to the NP effects.
We estimate the possible range of $\delta$ by varying the Wilson coefficients within the bounds discussed in Sec.~\ref{sec:WCconstraint}.
In this estimate we consider the single-operator NP scenarios of $S_L$, $S_R$, and $T$, together with the aligned scalar direction $C_{S_L}=C_{S_R}$.
As in Sec.~\ref{sec:SRviolation}, we scan $\alpha$ while imposing $\alpha+\beta=1$.

Figure~\ref{fig:Deldel} shows the resulting envelope of $\delta$ as a function of $\alpha$.
For each value of $\alpha$, the envelope is obtained by maximizing and minimizing $\delta$ over the NP scenarios described above.
The gray region denotes the range where at least one of the cancellation measures becomes $\epsilon_{kl}>1$.
For values of $\alpha$ for which $\epsilon_{kl}<1$ holds for all operator combinations considered here, the allowed violation remains below the percent level.
The narrowest envelope occurs around $\alpha \simeq 0.45$ yielding $|\delta| \approx 0.003$, close to the value of $\alpha_{TT}$ obtained from Eq.~\eqref{eq:KITSR}.
Thus, although some of $\delta_{kl}$ are sizable, the current bounds on the Wilson coefficients strongly limit the actual violation of the sum rule in the region where the cancellation is supported.

%%%%%%%%%%%%%%%%%%%%%%%%%%%%%%%%%%%%%%%%%%%%%%%%%
\begin{figure*}[!t]
\begin{center} 
\includegraphics[width=0.6\textwidth]{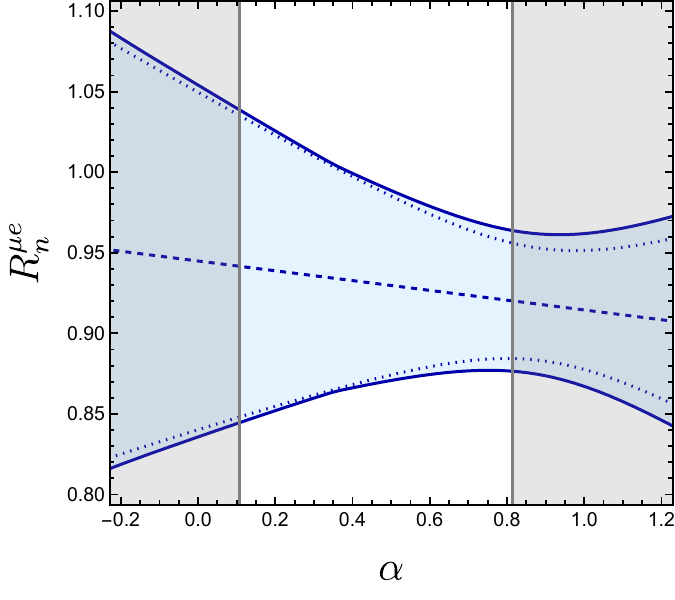}
\caption{
Sum rule prediction for $R^{\mu e}_n$ as a function of $\alpha$.
The blue shaded band shows the $1\,\sigma$ range from the experimental uncertainties in $R^{\mu e}_\pi$ and $R^{\mu e}_\rho$, including the allowed residual violation $\delta$.
The dotted contours correspond to the same range with $\delta=0$, and the dashed line is obtained from the central experimental inputs with $\delta=0$.
The gray region corresponds to the range where at least one of the cancellation measures satisfies $\epsilon_{kl}>1$.
} 
\label{fig:Rnemu}
\end{center}
\vspace{-3mm}
\end{figure*}
%%%%%%%%%%%%%%%%%%%%%%%%%%%%%%%%%%%%%%%%%%%%%%%%%
The preceding analysis shows that the residual violation allowed by the current bounds on the Wilson coefficients is at most at the percent level.
As will be seen below, this effect is smaller than the present experimental uncertainties in the mesonic inputs entering the sum-rule prediction.
We can therefore use the relation to predict the yet-unmeasured baryonic ratio $R_n^{\mu e}$ from the available mesonic inputs,
\begin{align}
    \label{eq:RnPrediction}
    R^{\mu e}_n
    =
    R^{\mu e,\rm SM}_n
    \left[
    \alpha \frac{R^{\mu e,\rm ex}_\pi}{R^{\mu e,\rm SM}_\pi}
    +
    \beta \frac{R^{\mu e,\rm ex}_\rho}{R^{\mu e,\rm SM}_\rho}
    +
    \delta
    \right],
\end{align}
where $R^{\mu e,\rm ex}_X$ is the experimental value of the corresponding ratio.
In addition, we obtained the SM value $R^{\mu e,\rm SM}_{n}=0.977$ using the form factor inputs collected in Appendix~\ref{app:FF}.
Its theoretical uncertainty is neglected, since it is expected to be smaller than the current experimental uncertainties.

The BESIII measurement of $R_\pi^{\mu e}$ yields \cite{BESIII:2018nzb}
\begin{align}
    R^{\mu e,\rm ex}_{\pi}
    =
    \frac{{\rm BR}(D^0\to\pi^-\ov{\mu}\nu)}
         {{\rm BR}(D^0\to\pi^-\ov{e}\nu)}
    =
    0.922\pm0.037\,.
\end{align}
At present, the statistical uncertainty is slightly larger than the systematic uncertainty, and a more precise measurement is expected in the future.
The SM prediction is given in Ref.~\cite{Riggio:2017zwh} as
\begin{align}
    R^{\mu e,\rm SM}_{\pi}=0.985\pm0.002\,.
\end{align}
This corresponds to a mild deviation of about $1.7\,\sigma$ between the experimental result and the SM prediction.

%%%%%%%%%%%%%%%%%%%%%%%%%%%%%%%%%%%%%%%%
\begin{table}[!t]
  \renewcommand{\arraystretch}{1.25}
  \centering
  \begin{tabular}{ccccccc} 
    $\alpha$ & $R^{\mu e}_{n}$ & Allowed range of $\delta$ & $ \mathrm{max} [\epsilon_{kl}]$   
    \\ \hline
%%%%%%%
    $0.25$ & $0.937 \pm 0.079$ & $[-0.0054,\,0.005
    2]$ &  $\epsilon_{V_LT}=0.67$   
    \\ \hline
%%%%%%%
   $0.45$ & $0.931 \pm 0.060$ &$[-0.0032,\,0.0027]$ & $\epsilon_{S_RS_L}=0.74$   
    \\ \hline
%%%%%%%
    $0.80$ & $0.920 \pm 0.036$ &  $[-0.0084,\,0.0057]$ & $\epsilon_{S_RS_L}=0.88$   
    \\ \hline
  \end{tabular}
  \caption{
  Predictions of $R^{\mu e}_{n}$ evaluated at $\delta=0$ for representative values of $\alpha$.
  Here, the uncertainties are from the propagated $1\sigma$ experimental uncertainties in $R_\pi^{\mu e}$ and $R_\rho^{\mu e}$. 
  The third column shows the allowed range of the residual sum rule violation $\delta$. 
  The last column shows the largest cancellation measure $\epsilon_{kl}$ and the corresponding operator combination.
  }
  \label{tab:RepRn}
\end{table}
%%%%%%%%%%%%%%%%%%%%%%%%%%%%%%%%%%%%
For $R_\rho^{\mu e}$, BESIII \cite{BESIII:2021pvy} and PDG \cite{PDG2024} report the branching fractions for $D^0\to\rho^-\ov{\mu}\nu$ and $D^0\to\rho^-\ov{e}\nu$ separately.
However, to our knowledge, no up-to-date direct determination of $R_\rho^{\mu e,\rm ex}$ is currently available.
BESIII \cite{BESIII:2021pvy} also provided an estimate of $R_\rho^{\mu e,\rm ex}$, which is consistent with the value obtained by combining the experimental branching fractions available at that time, without taking correlations among the uncertainties into account.
Combining the latest results \cite{PDG2024} without accounting for those correlations, we obtain
\begin{align}
    R^{\mu e,\rm ex}_\rho
    =
    \frac{{\rm BR}(D^0\to\rho^-\ov{\mu}\nu)}
         {{\rm BR}(D^0\to\rho^-\ov{e}\nu)}
    =
    0.925\pm0.102\,.
\end{align}
For the SM prediction, we are not aware of any study that directly evaluates $R_\rho^{\mu e,\rm SM}$ including its uncertainty.
We obtain the central value as
\begin{align}
    R^{\mu e,\rm SM}_{\rho}=0.956\,.
\end{align}
As can be seen in $R_\pi^{\mu e,\rm SM}$, the theoretical uncertainty in the ratio $R_H^{\mu e}$ is expected to be suppressed.
We therefore neglect the SM uncertainties in the present analysis.

Figure~\ref{fig:Rnemu} shows the sum-rule prediction for $R^{\mu e}_n$ as a function of $\alpha$.
The blue shaded region represents the $1\,\sigma$ uncertainty band obtained from the experimental uncertainties in $R_\pi^{\mu e}$ and $R_\rho^{\mu e}$, with the allowed shift due to $\delta$ included.
The blue dotted contours show the corresponding $1\,\sigma$ band with $\delta=0$.
In the gray region, at least one of the cancellation measures $\epsilon_{kl}$ exceeds unity, and the cancellation is therefore not efficient.
The small difference between the blue shaded region and the blue dotted contours shows that the residual sum rule violation is currently much smaller than the experimental uncertainties.
Thus, within the present experimental accuracy, the sum-rule prediction remains useful after including the residual violation, and its uncertainty is dominated by the mesonic input data.
For representative values of $\alpha$, Table~\ref{tab:RepRn} summarizes the predicted values of $R_n^{\mu e}$, the allowed ranges of the residual violation $\delta$, and the largest cancellation measure.
At $\alpha=0.8$, the two largest cancellation measures $\epsilon_{S_RS_L}$ and $\epsilon_{V_LT}$ are nearly degenerate.

Although the actual violation $\delta$ is minimized around $\alpha=0.4$--$0.45$, the smallest uncertainty in the estimate of $R_n^{\mu e}$ is obtained at a larger value of $\alpha$, around $\alpha=0.8$.
This behavior is driven by the current experimental inputs.
Since the uncertainty in $R_\rho^{\mu e}$ is much larger than that in $R_\pi^{\mu e}$, increasing $\alpha$ suppresses the contribution of $R_\rho^{\mu e}$ in Eq.~\eqref{eq:RnPrediction} and narrows the uncertainty band.
However, this region lies close to the gray region in Fig.~\ref{fig:Rnemu}, where the cancellation is not efficient.
Thus, the choice of $\alpha$ involves a tradeoff between reducing the propagated experimental uncertainty and maintaining an efficient cancellation in the sum rule.
With the present inputs, the resulting prediction has an uncertainty of about $4\%$, which provides a useful target for future measurements and consistency tests.
The prediction can be systematically updated as more precise measurements of $R_\pi^{\mu e}$ and $R_\rho^{\mu e}$ become available.

%%%%%%%%%%%%%%%%%%%%%%%%%%%%%%%%%%%%%%%%%%%
\section{Summary and discussion}
\label{sec:Summary}
%%%%%%%%%%%%%%%%%%%%%%%%%%%%%%%%%%%%%%%%%%%

Motivated by the sum rules in bottom semileptonic decays, we investigated whether an analogous relation can be constructed in charm hadron decays.
We focused on the $c\to d\ov\ell\nu$ transitions involving $D\to\pi\ov\ell\nu$, $D\to\rho\ov\ell\nu$, and $\Lambda_c\to n\ov\ell\nu$.
For these modes, we constructed a relation among the lepton-flavor universality ratios $R_H^{\mu e}$ and quantified its residual violation.
We found that the charm case exhibits a nontrivial cancellation of the residual violation.
Although this cancellation is less efficient compared with the $b\to c$ and $b\to u$ transitions, the scan over $\alpha$ shows that the charm relation remains useful over a broad range of $\alpha$.

We then included the current constraints on the NP contributions from $D^-\to\mu\ov\nu$ and high-$p_T$ searches.
Although some of the violations are sizable at the coefficient level, the allowed Wilson coefficients strongly restrict the actual residual violation $\delta$.
As a result, the actual violation is at most at the percent level and remains smaller than the present experimental uncertainties in the mesonic inputs.
Thus, the sum rule can still be used as a phenomenological consistency relation.

As an application, we used the relation to predict the yet-unmeasured ratio $R_n^{\mu e}$.
The present uncertainty is driven primarily by the experimental uncertainty in the mesonic inputs, especially in $R_\rho^{\mu e}$, and effects from the residual sum rule violation are negligible.
A larger value of $\alpha$ reduces the propagated uncertainty by suppressing the contribution from $R_\rho^{\mu e}$.
However, this region is close to where the cancellation becomes less efficient for some operator combinations.
The choice of $\alpha$ therefore involves a tradeoff between reducing the experimental uncertainty and maintaining the quality of the sum-rule cancellation.
With the current data, the resulting prediction has an uncertainty of about $4\%$, providing a useful target for future measurements of $\Lambda_c\to n\ov\ell\nu$.

In this paper, we assumed NP contributions only in the muon mode.
NP contributions to the electron mode could be analyzed in the same framework.
Since stringent high-$p_T$ constraints are also available \cite{Fuentes-Martin:2020lea}, one may expect a similarly predictive relation, although a dedicated analysis is needed to quantify this statement.
If NP effects are present simultaneously in the electron and muon modes, the lepton-flavor universality ratios may remain close to their SM values.
In such a case, sum rules relating branching fractions themselves could become more appropriate.

We also did not include form factor uncertainties in the main numerical analysis.
This limitation is especially relevant for $D\to\rho$ transitions, where the form factors enter the coefficients $a_{ij}^{\rho}$ and can potentially affect the cancellation pattern.
As a compact check, we repeated the analysis by changing only the $D\to\rho$ form factors, while keeping all the other inputs fixed.
Using the form factors including higher-twist contributions in Ref.~\cite{Wu:2006rd} instead of the leading-twist input used in the main analysis, we found that $R_{\rho}^{\mu e,\mathrm{SM}}$ changes only slightly, from $0.956$ to $0.953$.
This small change is expected because most of the form factor dependence cancels in the ratio of the muon and electron modes.
By contrast, some individual coefficients $a_{ij}^{\rho}$, especially the scalar coefficient $a_{SS}^{\rho}$, can differ by up to about $25\%$. 
Nevertheless, the values of $\alpha$ at which $\epsilon_{kl}$ becomes minimal change by at most about $5\%$, and the region satisfying $\epsilon_{kl}<1$ remains almost unchanged.
We also found that the cancellation quality is somewhat more sensitive to the $D\to\rho$ input in the large-$\alpha$ region.
However, the prediction for $R_n^{\mu e}$ is not significantly affected because the allowed NP Wilson coefficients are small.
This comparison should be regarded as a case study rather than a full form factor uncertainty analysis, and a more precise determination of the $D\to\rho$ form factors would be useful for a more quantitative assessment.

The present study gives a numerical first step toward semileptonic sum rules in heavy-to-light charm decays.
A fundamental understanding would be required to clarify whether the observed cancellations reflect an underlying structure or an accidental numerical pattern.
Future measurements of $R_\pi^{\mu e}$, $R_\rho^{\mu e}$, and eventually $R_n^{\mu e}$ will provide direct tests of this relation.

%%%%%%%%%%%%%%%%%%%%%%%%%%%%%%%%%%%%%%%%%
\section*{Acknowledgements}
%%%%%%%%%%%%%%%%%%%%%%%%%%%%%%%%%%%%%%%%%

The authors thank Hiroyasu Yonaha, Andreas Crivellin and Teppei Kitahara for the inspiring and encouraging discussions.
We also appreciate Wen-Feng Duan for providing a general formula for the $b\to u$ semi-baryonic decay.
%---------------------------------------------------------------------------
This work is supported by JSPS KAKENHI Grant Numbers 22K21347 [M.E. and S.I.], 24K07025 [S.M.], and 24K22879 [S.I.], 24K23939 [S.I.], and 25K17385 [S.I.].
The work of S.I. is also supported by Toyoaki scholarship foundation.
T.U. is supported by JST SPRING, Grant Number JPMJSP2125 and “THERS Make New Standards Program for the Next Generation Researchers”.

%%%%%%%%%%%%%%%%%%%%%%%%%%%%%%%%%%%%%
\appendix
%%%%%%%%%%%%%%%%%%%%%%%%%%%%%%%%%%%%%
\section{Form factor and hadronic amplitude}
\label{app:FF}
%%%%%%%%%%%%%%%%%%%%%%%%%%%%%%%%%%%%%
This appendix summarizes the form factors and helicity amplitudes used in the analysis.
We follow the form factor definition in Ref.~\cite{Meinel:2017ggx} as
\begin{align}
 \langle n(k) | \ov d \gamma_\mu c | \Lambda_c (p) \rangle 
 & = \ov u_{n}\biggl[F_0^{n}(m_\Lc-m_n)\frac{q_\mu}{q^2} \notag \\[0.5em]
 &~~~~~~~~+ F_+ \frac{m_\Lc+m_n}{Q_+} \left( q^\prime_\mu -(m_\Lc^2-m_n^2)\frac{q_\mu}{q^2} \right)\\ \nonumber
 &~~~~~~~~+ F_\perp \left(  \gamma_\mu-\frac{2m_n}{Q_+}p_\mu-\frac{2m_\Lc}{Q_+}k_\mu\right)\biggr] u_{\Lambda_c}\,,\\
 %%%
 \langle n(k) | \ov d \gamma_\mu\gamma_5 c | \Lambda_c (p) \rangle 
 & = - \ov u_{n} \gamma_5 \biggl[G_0(m_\Lc+m_n)\frac{q_\mu}{q^2} \notag \\[0.5em]
 &~~~~~~~~+ G_+ \frac{m_\Lc-m_n}{Q_-} \left( q^\prime_\mu -(m_\Lc^2-m_n^2)\frac{q_\mu}{q^2} \right)\\ \nonumber
 &~~~~~~~~+ G_\perp  \left(  \gamma_\mu+\frac{2m_n}{Q_-}p_\mu-\frac{2m_\Lc}{Q_-}k_\mu\right)\biggr] u_{\Lambda_c} \,,\\
 %%%
 \langle n(k) | \ov d c | \Lambda_c (p) \rangle 
 & = F_0^{n} \frac{m_\Lc-m_n}{m_c-m_d}  \ov u_{n} u_{\Lambda_c}\,,\\
 %%%
 \langle n(k) | \ov d \gamma_5 c | \Lambda_c (p) \rangle 
 & = G_0 \frac{m_\Lc+m_n}{m_c+m_d} \ov u_{n}  \gamma_5 u_{\Lambda_c} \,,\\
%%%
 \langle n(k) | \ov d \sigma_{\mu\nu} c | \Lambda_c (p) \rangle 
 & = \ov u_{n} \biggl[ -2i H_+  \frac{p_\mu k_\nu- k_\mu p_\nu}{Q_+}\\ \nonumber
 & -i H_\perp\left( \frac{m_\Lc+m_n}{q^2} (q_\mu \gamma_\nu - \gamma_\mu q_\nu)-2\left( \frac{1}{q^2}+\frac{1}{Q_+}\right) \left( p_\mu k_\nu- k_\mu p_\nu \right)  \right)\\ \nonumber
 & + \tilde H_+ \left( \sigma_{\mu\nu}+\frac{2i}{Q_-}\left( m_\Lc (k_\mu \gamma_\nu - \gamma_\mu k_\nu) -m_n (p_\mu \gamma_\nu - \gamma_\mu p_\nu)+ p_\mu k_\nu- k_\mu p_\nu \right)\right) \\ \nonumber
 & -i \tilde H_\perp \frac{m_\Lc-m_n}{q^2 Q_-} \biggl( (m_\Lc^2-m_n^2-q^2) (\gamma_\mu p_\nu -p_\mu \gamma_\nu)\\ \nonumber
 &- (m_\Lc^2-m_n^2+q^2) (\gamma_\mu k_\nu -k_\mu \gamma_\nu)+(m_\Lc-m_n)(p_\mu k_\nu-k_\mu p_\nu) \biggr) \biggr] u_{\Lambda_c}\, ,
\end{align} 
with $q=p-k$, $q^\prime=p+k$, and $Q_\pm=(m_\Lc\pm m_n)^{2}-q^2$.

For the $D\to \pi$ transition, we use the form factor definitions \cite{Sakaki:2013bfa}
\begin{align}
\langle \pi (k) | \ov d \gamma_\mu c | D (p) \rangle 
 & = F_1 \left( q^\prime_\mu -\frac{m_D^2-m_\pi^2}{q^2}q_\mu\right) +F_0^{\pi} \frac{m_D^2-m_\pi^2}{q^2} q_\mu\, , \\ 
%%%%%
\langle \pi (k) | \ov d  c | D (p) \rangle 
 & = F_0^{\pi} \frac{m_D^2-m_\pi^2}{m_c-m_d}\, , \\ 
%%%%%
\langle \pi (k) | \ov d \sigma_{\mu\nu}  c | D (p) \rangle 
 & = -i\frac{2F_T}{m_D+m_\pi} \left(p_\mu k_\nu-k_\mu p_\nu \right)\, , 
\end{align}
and the $D\to \rho$ transition as
\begin{align}
\langle \rho (k,\,\epsilon) | \ov d \gamma_\mu c | D (p) \rangle 
 & = -i\frac{2V}{m_D+m_\rho}  \epsilon_{\mu\nu\alpha\beta}\, \epsilon^{\nu*} p^\alpha k^\beta\, , \\ 
 %%%%
\langle \rho (k,\,\epsilon) | \ov d \gamma_\mu \gamma_5 c | D (p) \rangle 
 & = A_1 (m_D+m_\rho) \epsilon^*_\mu - A_2 \frac{\epsilon^*\cdot q}{m_D+m_\rho} q^\prime_\mu\, \\ \nonumber
 &~~~~-\left(A_3-A_0\right)\frac{2m_\rho\, \epsilon^*\cdot q}{q^2} q_\mu\, ,  \\
 %%%%
 \langle \rho (k,\,\epsilon) | \ov d \gamma_5 c | D (p) \rangle 
 & = -A_0\frac{2m_\rho\,\epsilon^*\cdot q}{m_c+m_d}\, , \\ 
 %%%%%%%
 \langle \rho (k,\,\epsilon) | \ov d \sigma_{\mu\nu}c | D (p) \rangle 
 & = \epsilon_{\mu\nu\alpha\beta} \biggl\{ -T_1 \epsilon^{*\alpha}q^{\prime\beta}+ \left( T_1-T_2\right) \frac{m_D^2-m_\rho^2}{q^2} \epsilon^{*\alpha}q^{\beta} \\ \nonumber
 &~~~~~~~~~~~+\left(T_1-T_2-T_3\frac{q^2}{m_D^2-m_\rho^2}\right) \frac{2\,\epsilon^*\cdot q}{q^2} p^\alpha k^\beta\biggr\}\, ,  
\end{align}
with the convention of $\ov{d}\sigma^{\mu\nu}\gamma_5 c=-(i/2)\epsilon^{\mu\nu\alpha\beta} \, \ov{d} \sigma_{\alpha\beta} c$ and $\epsilon_{0123}=-\epsilon^{0123}=1$.
We also define
\begin{align}
\label{eq:defA3}
A_3=A_1\frac{m_D+m_\rho}{2m_\rho}-A_2\frac{m_D-m_\rho}{2m_\rho}\,. 
\end{align}

The helicity amplitudes are given for the mesonic decays as 
\begin{align}
    H^\pi_{V,0}=&\sqrt{\frac{\lambda_\pi}{q^2}}F_1\,,\\
    H^\pi_{V,t}=&\frac{m_D^2-m_\pi^2}{\sqrt{q^2}}F_0^{\pi},\\
    H^\pi_{S}=&\frac{m_D^2-m_\pi^2}{m_c-m_d}F_0^{\pi}\,,\\
    H^\pi_{T}=&-\frac{\sqrt{\lambda_\pi}}{m_D+m_\pi}F_T\,,\\
    H^\rho_{V,\pm}=&(m_D+m_\rho)A_1\mp\frac{\sqrt{\lambda_\rho}}{m_D+m_\rho}V\,,\\
    H^\rho_{V,0}=&\frac{m_D+m_\rho}{2m_\rho\sqrt{q^2}}\Bigl[-(m_D^2-m_\rho^2-q^2)A_1+\frac{\lambda_\rho}{(m_D+m_\rho)^2}A_2\Bigr]\,,\\
    H^\rho_{V,t}=&-\sqrt{\frac{\lambda_\rho}{q^2}}A_0\,,\\
    H^\rho_{P}=&-\frac{\sqrt{\lambda_\rho}}{m_c+m_d}A_0\,,\\
    H^\rho_{T,\pm}=&\frac{1}{\sqrt{q^2}}\bigl[\pm(m_D^2-m_\rho^2)T_2+\sqrt{\lambda_\rho}T_1\bigr]\,,\\
    H^\rho_{T,0}=&\frac{1}{2m_\rho}\Bigl[-(m_D^2+3m_\rho^2-q^2)T_2+\frac{\lambda_\rho}{m_D^2-m_\rho^2}T_3\Bigr]\,.
\end{align}
For the baryonic decay, the helicity amplitudes are given by 
\begin{align}
H^n_{V,1\pm}=&-\sqrt{2Q_-}F_\perp \pm \sqrt{2Q_+}G_\perp\,,\\
H^n_{V,0\pm}=&\sqrt{\frac{Q_-}{q^2}}(m_{\Lambda_{c}}+m_n)F_+ \mp \sqrt{\frac{Q_+}{q^2}}(m_{\Lambda_{c}}-m_n)G_+\,,\\
H^n_{V,t\pm}=&\sqrt{\frac{Q_+}{q^2}}(m_{\Lambda_{c}}-m_n)F_0^n \mp \sqrt{\frac{Q_-}{q^2}}(m_{\Lambda_{c}}+m_n)G_0\,,\\
H^n_{S}=&\frac{m_{\Lambda_{c}}-m_n}{m_c-m_d}\sqrt{Q_+}F_0^n\,,\\
H^n_{P}=&\frac{m_{\Lambda_{c}}+m_n}{m_c+m_d}\sqrt{Q_-}G_0\,,\\
H^n_{T,1\pm}=&-\sqrt{\frac{2}{q^2}}\bigl[ (m_{\Lambda_{c}}+m_n)\sqrt{Q_-}H_\perp \pm (m_{\Lambda_{c}}-m_n)\sqrt{Q_+} \tilde H_\perp \bigr]\,,\\
H^n_{T,0\pm}=&\sqrt{Q_-}H_+ \pm \sqrt{Q_+} \tilde H_+\,.
\end{align}

We next summarize the form factor parameterizations and numerical inputs used in this work.
For the $D\to \pi$ form factors, we use the result of LQCD \cite{Lubicz:2017syv,Lubicz:2018rfs}.
Form factors are expressed as
\begin{align}
F_{\pi}(q^2)=
\frac{F_{\pi}(0)+ c^{\pi} (z^{\pi} - z^{\pi}_0)\left( 1 + \frac{z^{\pi} + z^{\pi}_0}{2} \right)}
{1 - P_{\pi} \, q^2} \, ,
\end{align}
where $ F_{\pi}=\left\{F_0^{\pi},\, F_1,\, F_T\right\} $, with $c^{\pi}=\left\{c_0,\, c_1,\, c_T\right\} $ and $ P_{\pi}=\left\{P_S,\, P_V,\, P_T\right\} $, respectively.
Here, we define $ z^{\pi} $ as
\begin{align}
\label{eq:func-z}
z^{\pi}(q^2)=
\frac{\sqrt{t_+ - q^2} - \sqrt{t_+ - t_0}}
{\sqrt{t_+ - q^2} + \sqrt{t_+ - t_0}} \, ,
\end{align}
with $ t_+ = (m_D + m_{\pi})^2 $, $ t_0 = (m_D + m_{\pi})\left( \sqrt{m_D} - \sqrt{m_{\pi}} \right)^2 $, and $ z^{\pi}_0=z^{\pi}(0) $.
The numerical values are 
\begin{align}
\left\{ F_0^{\pi}, \, F_1, \, F_T \right\}
&=
\left\{ 0.6117, \, 0.6117, \, 0.5063 \right\}\, ,\\
\left\{ c_0, \, c_1, \, c_T \right\}
&=
\left\{ -1.188, \, -1.985, \, -1.10 \right\}\, ,\\
\left\{ P_S, \, P_V, \, P_T \right\}
&=
\left\{ 0.0342, \, 0.1314, \, 0.1461 \right\}\,\rm{GeV}^{-2} .
\end{align}

For the $D \to \rho$ form factors, we adopt the LCSR results of Ref.~\cite{Wu:2006rd}.
The vector-current form factors can be written as
\begin{align}
F_{\rho}(q^2) = \frac{F_{\rho}(0)}{1 - a_{\rho} \, \frac{q^2}{m_D^2} + b_{\rho} \left( \frac{q^2}{m_D^2} \right)^2}\,,
\end{align}
where $ F_{\rho}=\left\{V,\,A_1,\,A_2,\,A'_3 \right\}$, with $ a_{\rho}=\left\{a_V,\,a_1,\,a_2,\,a_3\right\} $ and $ b_{\rho}=\left\{b_V,\,b_1,\,b_2,\,b_3\right\} $, respectively.
Here, $ A_3(q^2) $ is obtained from Eq.~(\ref{eq:defA3}), while $ A_0(q^2) $ is determined from
\begin{align}
\label{eq:defA0}
A_0=A_3-\frac{q^2}{2m_{\rho}(m_D+m_{\rho})} A'_3\,. 
\end{align}
The numerical values of the parameters used are\footnote{
This corresponds to the leading-twist scenario in Ref.~\cite{Wu:2006rd}.
The higher-twist scenario was found to receive larger higher-order corrections in the heavy-quark expansion, and we therefore use the leading-twist case.
}
\begin{align}
\left\{V(0), \, A_1(0), \, A_2(0), \, A'_3(0)\right\}
&=
\left\{0.735, \, 0.590, \, 0.528, \, -0.528\right\}\, ,\\
\left\{a_V,\,a_1,\,a_2,\,a_3\right\}
&=
\left\{0.48, \, 0.44, \, 0.91, \, 0.91\right\}\, ,\\
\left\{b_V,\,b_1,\,b_2,\,b_3\right\}
&=
\left\{2.25, \, 0.20, \, -1.01, \, -1.01\right\}\, . 
\end{align}
For the tensor form factors, no direct LCSR or lattice determination is available.
We therefore follow Ref.~\cite{Wu:2006rd} and use relations based on heavy-quark symmetry,
\begin{align}
T_1=&- \frac{m_D^2-m_\rho^2+q^2}{2m_D(m_D+m_\rho)}V -\frac{m_D+m_\rho}{2m_D} A_1 \, ,\\ 
T_2=&\frac{2}{m_D^2-m_\rho^2} \left[ \frac{\left(m_D-y \right)\left(m_D+m_\rho\right)}{2}A_1+\frac{m_D\left(y^2-m_\rho^2 \right)}{m_D+m_\rho}V \right] \, ,\\ 
T_3=&-\frac{m_D+m_\rho}{2m_D}A_1 +\frac{m_D-m_\rho}{2m_D}\left(A_2-A'_3\right)+\frac{m_D^2+3m_\rho^2-q^2}{2m_D\left(m_D+m_\rho\right)}V \, ,
\end{align}
where $y=(m_D^2+m_\rho^2-q^2)/(2m_D)$.
This treatment relies on the relations under the heavy-quark symmetry for the tensor contribution.
A direct determination of the $D\to\rho$ form factors, including the tensor form factors, would be needed for a more quantitative test of the charm sum rule.\footnote{
Recent LCSR studies indicate that finite-width and nonresonant-background effects may be relevant for precision studies \cite{Fu:2018yin,Lin:2025cmn}.
}
A lattice calculation treating the vector meson as a resonance has recently been demonstrated for $B\to\rho$ \cite{Leskovec:2025gsw}, and a corresponding calculation for charm decays would be particularly valuable.

For the $ \Lambda_c \to n $ form factors, we use the LQCD result \cite{Meinel:2017ggx}.
They can be written as
\begin{align}
\label{eq:FFfunc-n}
F_{n}(q^2) =
\frac{1}{1 - \frac{q^2}{(m_f^{\mathrm{pole}})^2}}\sum_{n=0}^{2} a_n^f [z^{\Lambda_c}(q^2)]^n\,,
\end{align}
where the form factors $F_n$, the coefficients $a_n^f$, and the pole masses $m_f^{\mathrm{pole}}$ are
\begin{align}
F_n &= \left\{F_0^n,\,F_+,\,F_\perp,\,G_0,\,G_+,\, G_\perp,\,H_\perp,\,\tilde{H}_\perp,\,H_+,\,\tilde{H}_+\right\}, \\
a_n^f &= \left\{a_n^{f_0},\,a_n^{f_+},\,a_n^{f_\perp},a_n^{g_0},\,a_n^{g_+},\,a_n^{g_\perp},\, a_n^{h_\perp},\,a_n^{\tilde{h}_\perp},\, a_n^{h_+},\,a_n^{\tilde{h}_+}\right\}, \\
m_f^{\mathrm{pole}} &= \left\{m_{f_0}^{\mathrm{pole}},\,m_{f_+}^{\mathrm{pole}},\,m_{f_\perp}^{\mathrm{pole}},\,m_{g_0}^{\mathrm{pole}},\,m_{g_+}^{\mathrm{pole}},\,m_{g_\perp}^{\mathrm{pole}}, m_{h_\perp}^{\mathrm{pole}},\,m_{\tilde{h}_\perp}^{\mathrm{pole}},\,m_{h_+}^{\mathrm{pole}},\,m_{\tilde{h}_+}^{\mathrm{pole}}\right\}.
\end{align}
The variable $z^{\Lambda_c}$ is defined as in Eq.~\eqref{eq:func-z}, with $t_0$ replaced by $(m_{\Lambda_c}-m_n)^2$ and $t_+$ remains unchanged accounting for the $D\pi$ threshold.
The numerical values of these parameters are
\begin{align}
\left\{a_0^{f_0}, \, a_1^{f_0}, \, a_2^{f_0}, \, a_0^{f_+}, \, a_1^{f_+}, \, a_2^{f_+}\right\}
=&
\left\{0.84, \, -2.57, \, 9.87, \, 0.83, \, -2.33, \, 8.41\right\}\,,\\
\left\{a_0^{f_\perp}, \, a_1^{f_\perp}, \, a_2^{f_\perp}, \, a_0^{g_0}, \, a_1^{g_0}, \, a_2^{g_0}\right\}
=&
\left\{1.36, \, -1.70, \, 0.71, \, 0.73, \, -0.97, \, 0.83\right\}\,,\\
\left\{a_0^{g_+}, \, a_1^{g_+}, \, a_2^{g_+}, \, a_0^{g_\perp}, \, a_1^{g_\perp}, \, a_2^{g_\perp}\right\}
=&
\left\{0.69, \, -0.90, \, 2.25, \, 0.69, \, -0.68, \, 0.70\right\}\,,\\
\left\{a_0^{h_\perp}, \, a_1^{h_\perp}, \, a_2^{h_\perp}, \, a_0^{\tilde{h}_\perp}, \, a_1^{\tilde{h}_\perp}, \, a_2^{\tilde{h}_\perp}\right\}
=&
\left\{0.63, \, -1.04, \, 1.42, \, 0.63, \, -1.39, \, 4.22\right\}\,, \\
\left\{a_0^{h_+}, \, a_1^{h_+}, \, a_2^{h_+}, \, a_0^{\tilde{h}_+}, \, a_1^{\tilde{h}_+}, \, a_2^{\tilde{h}_+}\right\}
=&
\left\{1.11, \, -0.69, \, -2.84, \, 0.63, \, -1.19, \, 3.73\right\}\,,\\
\label{eq:mpole}
m_{f_0}^{\mathrm{pole}}=2.351\,\rm{GeV},\,\,\,\,\,\,  
m_{f_+}^{\mathrm{pole}}= &\,\,m_{f_\perp}^{\mathrm{pole}}= m_{h_\perp}^{\mathrm{pole}}= m_{h_+}^{\mathrm{pole}}=2.010\,\rm{GeV},\\
m_{g_0}^{\mathrm{pole}}=1.870\,\rm{GeV},\,\,\,\,\,\,
m_{g_+}^{\mathrm{pole}}= &\,\,m_{g_\perp}^{\mathrm{pole}}= m_{\tilde{h}_\perp}^{\mathrm{pole}}= m_{\tilde{h}_+}^{\mathrm{pole}}=2.423\,\rm{GeV}.
\end{align}
From Eq.~(\ref{eq:FFfunc-n}), one finds that form factors develop a pole for $q^2_{\mathrm{max}} > (m_f^{\mathrm{pole}})^2$.

%%%%%%%%%%%%%%%%%%%%%%%%%%%%%
\section{Numerical formulae for bottom-hadron decays}
\label{app:bulnu}
%%%%%%%%%%%%%%%%%%%%%%%%%%%%%

This appendix collects the normalized branching fractions for the bottom transitions used for comparison in Sec.~\ref{sec:SRviolation}.
We define
\begin{align}
{\rm BR}^{q\mu}_X = {\rm BR}(H_b\to X\,\mu\ov\nu)\,.
\end{align}
The Wilson coefficients $C_i^{q\mu}$ are defined analogously to those in Eq.~\eqref{eq:Hamiltonian} with the final state $\mu\ov\nu$.
For $b\to c\,\mu\ov\nu$ transitions, we obtain
\begin{align}
    \frac{{\rm{BR}}_D^{b\mu}}{{\rm{BR}}_D^{b\mu,{\rm SM}}}& =  |1+C^{c\mu}_{V_L}|^2 
  +0.15\,{\rm{Re}}[ (1 +C^{c\mu}_{V_L}) C_{S_R}^{c\mu*}]
  +0.15\,{\rm{Re}}[ (1 +C^{c\mu}_{V_L}) C_{S_L}^{c\mu*} ]
\nonumber \\[0.5em]
&  + 2.11\, {\rm{Re}}[C^{c\mu}_{S_L} C_{S_R}^{c\mu*} ]
  +1.05\, (|C^{c\mu}_{S_L}|^2 + |C^{c\mu}_{S_R}|^2)
\nonumber \\[0.5em]
& +0.18\,{\rm{Re}}[ (1 +C^{c\mu}_{V_L})  C_T^{c\mu*}]  +0.64\, |C^{c\mu}_T|^2\,,\\
%%%%%
\frac{{\rm{BR}}_{D^*}^{b\mu}}{{\rm{BR}}_{D^*}^{b\mu,{\rm SM}}}& =  |1+C^{c\mu}_{V_L}|^2 
  +0.020\,{\rm{Re}}[ (1 +C^{c\mu}_{V_L}) C_{S_R}^{c\mu*}]
  -0.020\,{\rm{Re}}[ (1 +C^{c\mu}_{V_L}) C_{S_L}^{c\mu*} ]
\nonumber \\[0.5em]
&  -0.12\, {\rm{Re}}[C^{c\mu}_{S_L} C_{S_R}^{c\mu*} ]
 +0.060 \, (|C^{c\mu}_{S_L}|^2 + |C^{c\mu}_{S_R}|^2) 
\nonumber \\[0.5em]
& -0.47\,{\rm{Re}}[ (1 +C^{c\mu}_{V_L})  C_T^{c\mu*}] +  15.2\, |C^{c\mu}_T|^2\,,\\
%%%%%
\frac{{\rm{BR}}_{\Lambda_c}^{b\mu}}{{\rm{BR}}_{\Lambda_c}^{b\mu,{\rm SM}}}& =  |1+C^{c\mu}_{V_L}|^2 
  +0.050\,{\rm{Re}}[ (1 +C^{c\mu}_{V_L}) C_{S_R}^{c\mu*}]
  +0.024\,{\rm{Re}}[ (1 +C^{c\mu}_{V_L}) C_{S_L}^{c\mu*} ]
\nonumber \\[0.5em]
&  +0.48 \, {\rm{Re}}[C^{c\mu}_{S_L} C_{S_R}^{c\mu*} ]
  +0.33\, (|C^{c\mu}_{S_L}|^2 + |C^{c\mu}_{S_R}|^2) 
\nonumber \\[0.5em]
& -0.27 \,{\rm{Re}}[ (1 +C^{c\mu}_{V_L})  C_T^{c\mu*}] 
+11.6\, |C^{c\mu}_T|^2\,.
\end{align}
For $b\to u\mu\ov\nu$ transitions, we obtain
\begin{align}
    \frac{{\rm{BR}}_\pi^{b\mu}}{{\rm{BR}}_\pi^{b\mu,{\rm SM}}}& =  |1+C^{u\mu}_{V_L}|^2 
  +0.10\,{\rm{Re}}[ (1 +C^{u\mu}_{V_L}) C_{S_R}^{u\mu*}]
  +0.10\,{\rm{Re}}[ (1 +C^{u\mu}_{V_L}) C_{S_L}^{u\mu*} ]
\nonumber \\[0.5em]
&  + 3.32\, {\rm{Re}}[C^{u\mu}_{S_L} C_{S_R}^{u\mu*} ]
  +1.66\, (|C^{u\mu}_{S_L}|^2 + |C^{u\mu}_{S_R}|^2)
\nonumber \\[0.5em]
& +0.24\,{\rm{Re}}[ (1 +C^{u\mu}_{V_L})
C_T^{u\mu*}] + 3.35\, |C^{u\mu}_T|^2\,,\\
%%%%%
\frac{{\rm{BR}}_{\rho}^{b\mu}}{{\rm{BR}}_{\rho}^{b\mu,{\rm SM}}}& =  |1+C^{u\mu}_{V_L}|^2 
  +0.036\,{\rm{Re}}[ (1 +C^{u\mu}_{V_L}) C_{S_R}^{u\mu*}]
  -0.036\,{\rm{Re}}[ (1 +C^{u\mu}_{V_L}) C_{S_L}^{u\mu*} ]
\nonumber \\[0.5em]
&  -0.64\, {\rm{Re}}[C^{u\mu}_{S_L} C_{S_R}^{u\mu*} ]
 +0.32 \, (|C^{u\mu}_{S_L}|^2 + |C^{u\mu}_{S_R}|^2) 
\nonumber \\[0.5em]
& -0.19\,{\rm{Re}}[ (1 +C^{u\mu}_{V_L})  C_T^{u\mu*}] +  12.0\, |C^{u\mu}_T|^2\,,\\
%%%%%
\frac{{\rm{BR}}_{p}^{b\mu}}{{\rm{BR}}_{p}^{b\mu,{\rm SM}}}& =  |1+C^{u\mu}_{V_L}|^2 
  +0.048\,{\rm{Re}}[ (1 +C^{u\mu}_{V_L}) C_{S_R}^{u\mu*}]
  +0.0082\,{\rm{Re}}[ (1 +C^{u\mu}_{V_L}) C_{S_L}^{u\mu*} ]
\nonumber \\[0.5em]
&  + 0.31\, {\rm{Re}}[C^{u\mu}_{S_L} C_{S_R}^{u\mu*} ]
  +0.61\, (|C^{u\mu}_{S_L}|^2 + |C^{u\mu}_{S_R}|^2) 
\nonumber \\[0.5em]
& -0.019 \,{\rm{Re}}[ (1 +C^{u\mu}_{V_L})  C_T^{u\mu*}] 
+8.62\, |C^{u\mu}_T|^2\,.
\end{align}

%%%%%%%%%%%%%%%%%%%%%%%%%%%%%%%%%%%%%%%%%%%%
\section{$\delta_{kl}$ and $\epsilon_{kl}$ for bottom-hadron decays}
\label{app:al_vari}
%%%%%%%%%%%%%%%%%%%%%%%%%%%%%%%%%%%%%%%%%%%%

%%%%%%%%%%%%%%%%%%%%%%%%%%%%%%%%%%%%%%%%%%%%%%%%%
\begin{figure*}[t]
\begin{center}
\includegraphics[width=0.455\textwidth]{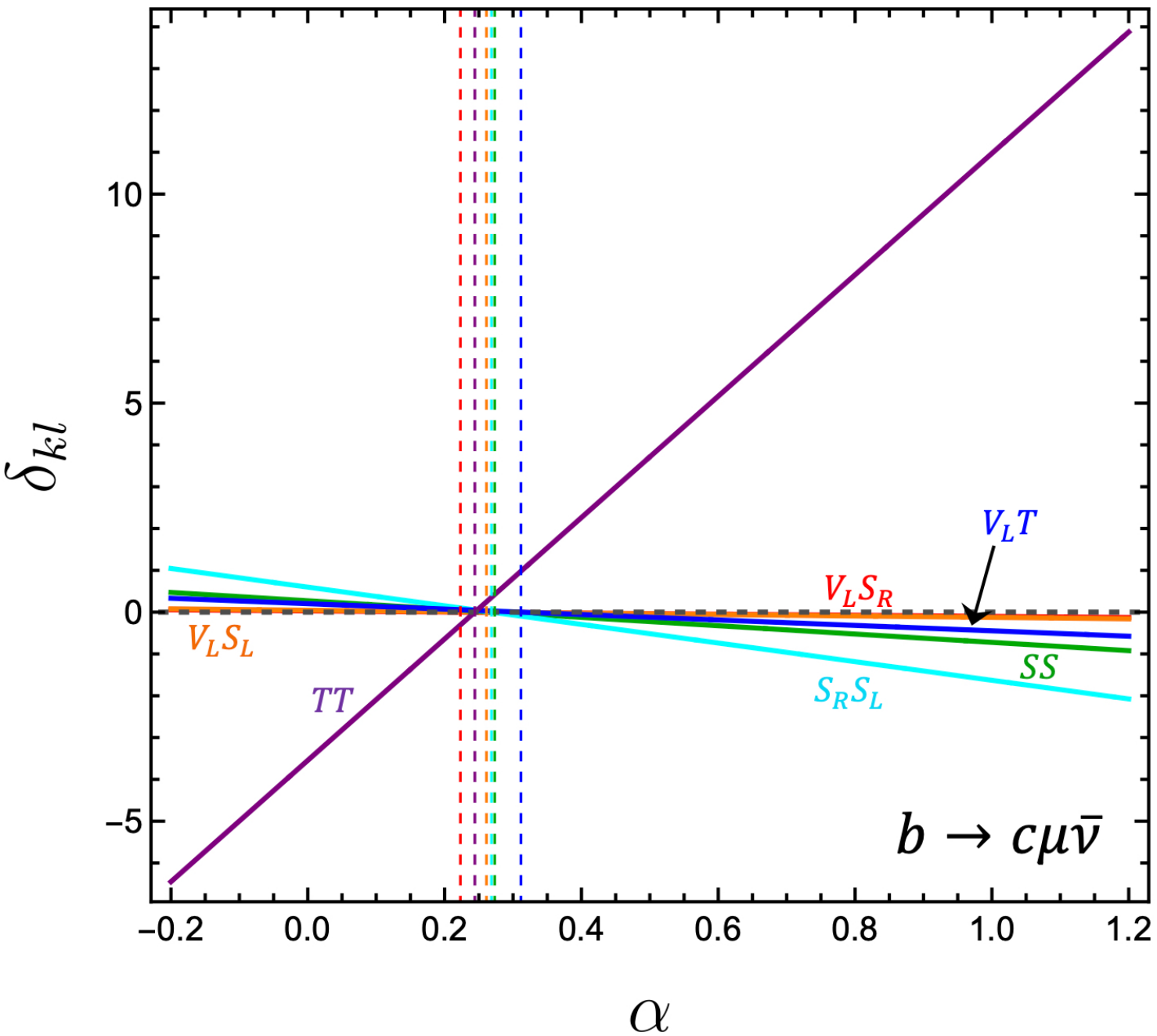}~~~
\includegraphics[width=0.45\textwidth]{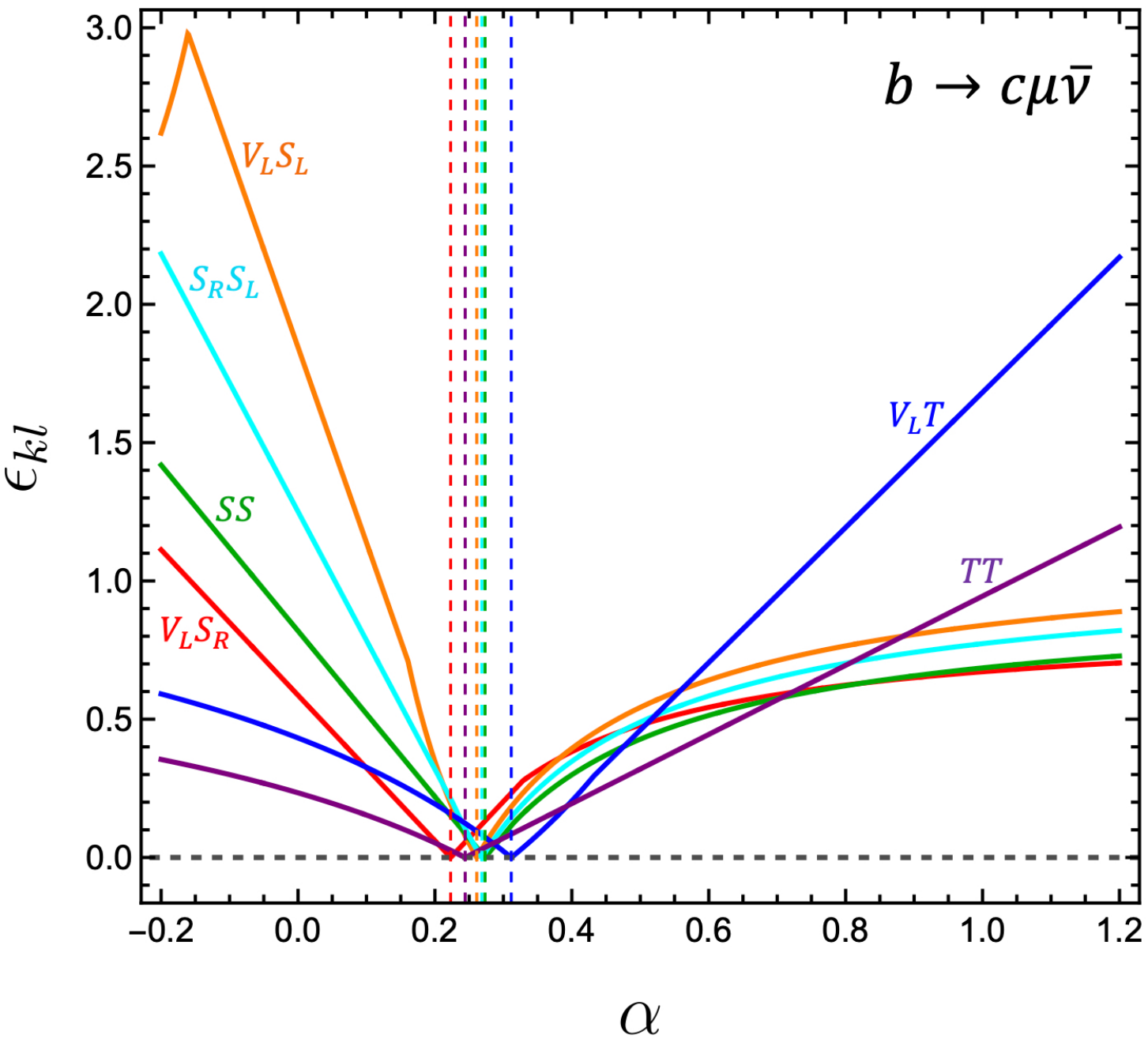}\\\vspace{5mm}
\includegraphics[width=0.45\textwidth]{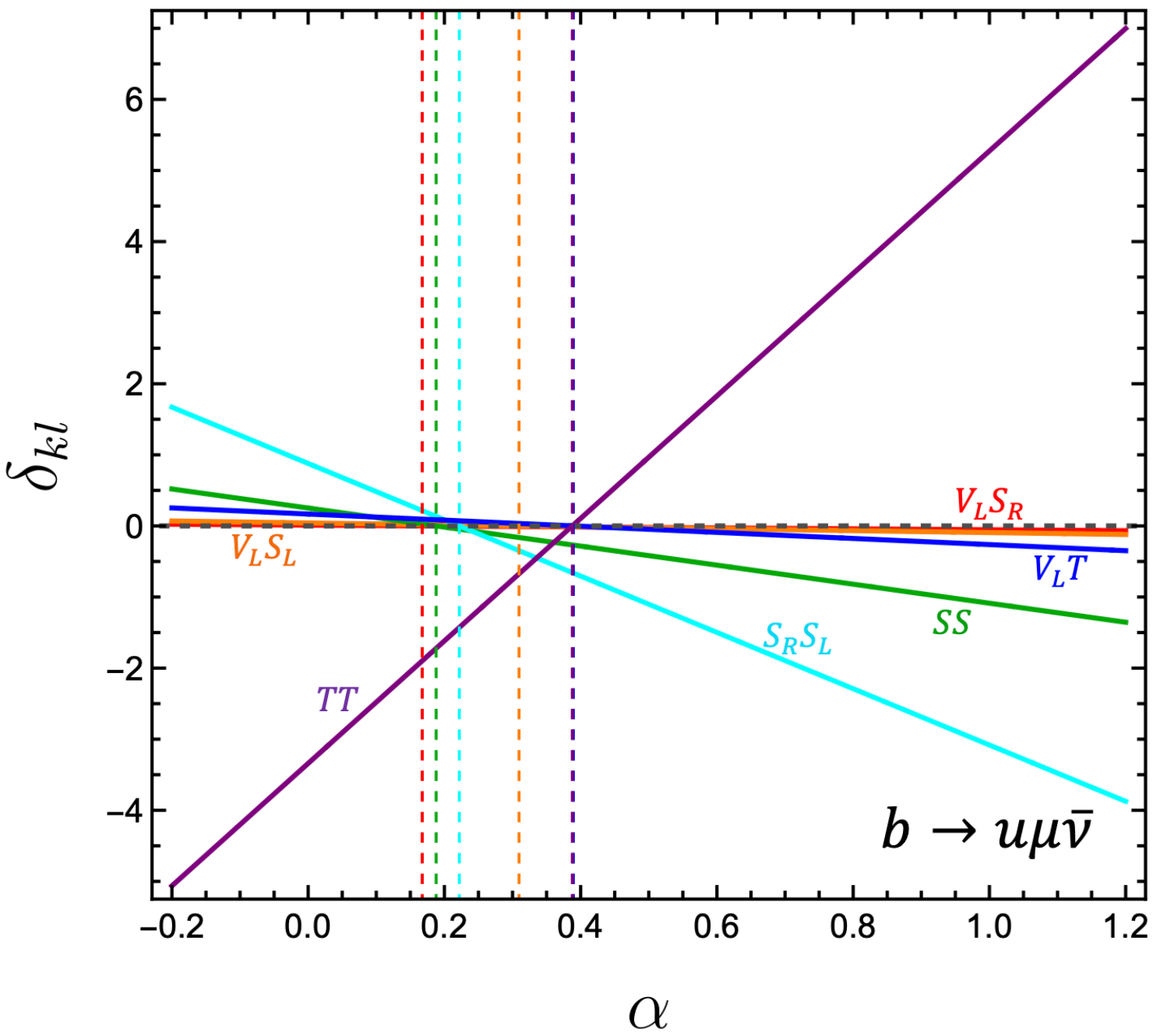}~~~
\includegraphics[width=0.45\textwidth]{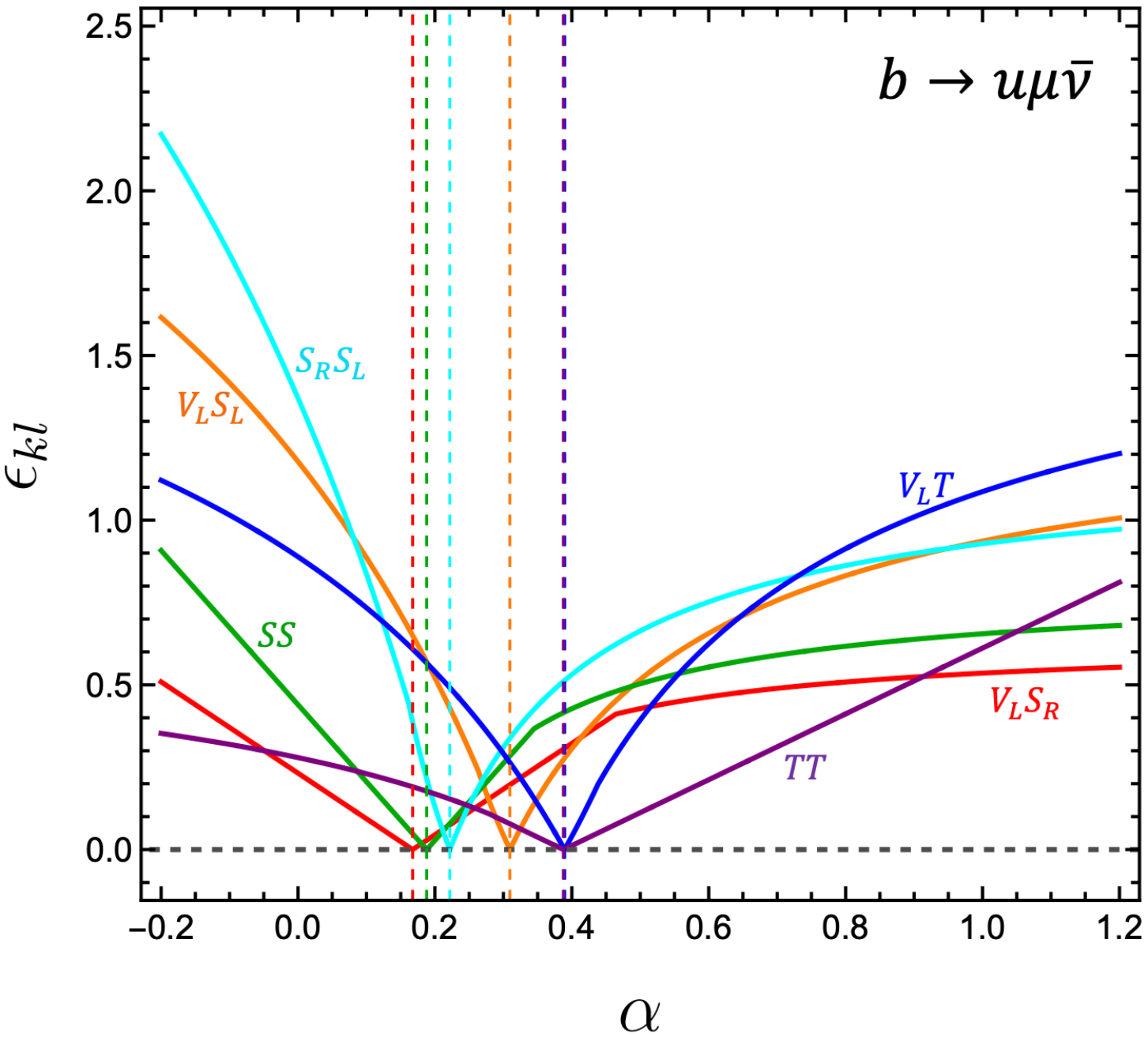}
\caption{
   The $\alpha$ dependence of $\delta_{kl}$ and $\epsilon_{kl}$ for $b\to c\mu\ov\nu$ (upper) and $b\to u\mu\ov\nu$ (lower).
   The left and right panels show $\delta_{kl}$ and $\epsilon_{kl}$, respectively.
   The vertical dashed lines indicate the values of $\alpha_{mn}$.
} 
\label{fig:deleps_alpha}
\end{center}
\vspace{-3mm}
\end{figure*}
%%%%%%%%%%%%%%%%%%%%%%%%%%%%%%%%%%%%%%%%%%%%%%%%%

Figure~\ref{fig:deleps_alpha} shows the $\alpha$ dependence of $\delta_{kl}$ and $\epsilon_{kl}$ for $b\to c\mu\ov\nu$ and $b\to u\mu\ov\nu$.
The values of $\alpha_{mn}$ obtained from Eq.~\eqref{eq:KITSR}, where each corresponding $\delta_{mn}$ vanishes, are also indicated.
Compared with the $c\to d$ case shown in Fig.~\ref{fig:deleps_alpha_cd}, the values of $\alpha_{mn}$ are more closely aligned in the bottom-hadron decays.
The spread is about $0.1$ for $b\to c\mu\ov\nu$ and about $0.2$ for $b\to u\mu\ov\nu$.
This alignment allows several operator combinations to be suppressed simultaneously for a common value of $\alpha$.
In particular, the $b\to c$ case exhibits efficient suppression around $\alpha \simeq 1/4$, as expected from the small-velocity limit.
The phenomenological implications of the residual violation also depend on the experimentally allowed ranges of the Wilson coefficients, which are constrained by $B$-meson decays and by high-$p_T$ searches at the LHC (see, for instance, Refs.~\cite{Iguro:2020keo,Iguro:2020cpg,Leljak:2023gna}).

%%%%%%%%%%%%%%%%%%%%%%%%%%%%%%%%%%%%%%%%%%%%%%%%%
\begin{figure*}[t]
\begin{center}
\includegraphics[width=0.45\textwidth]{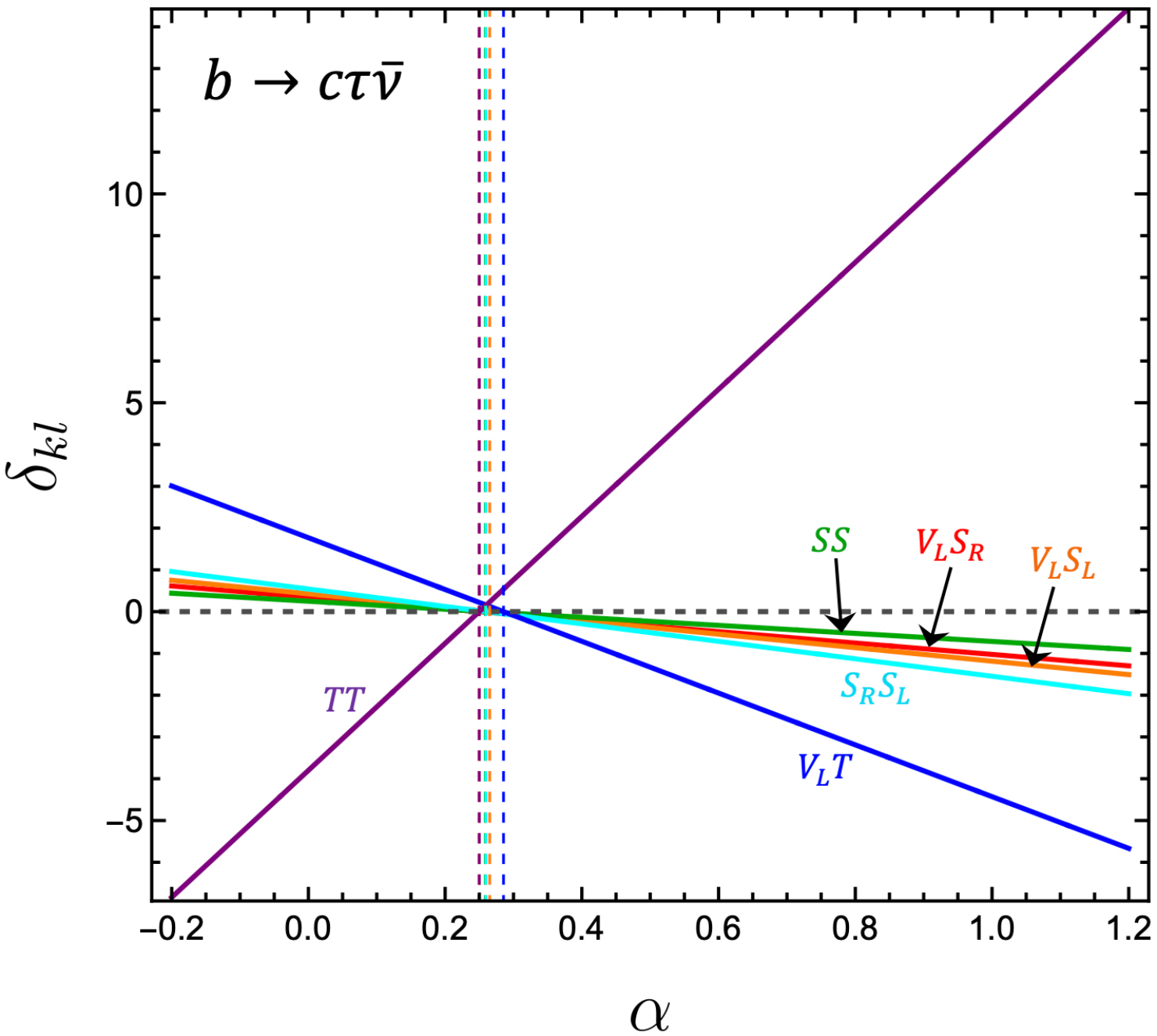}~~~
\includegraphics[width=0.449\textwidth]{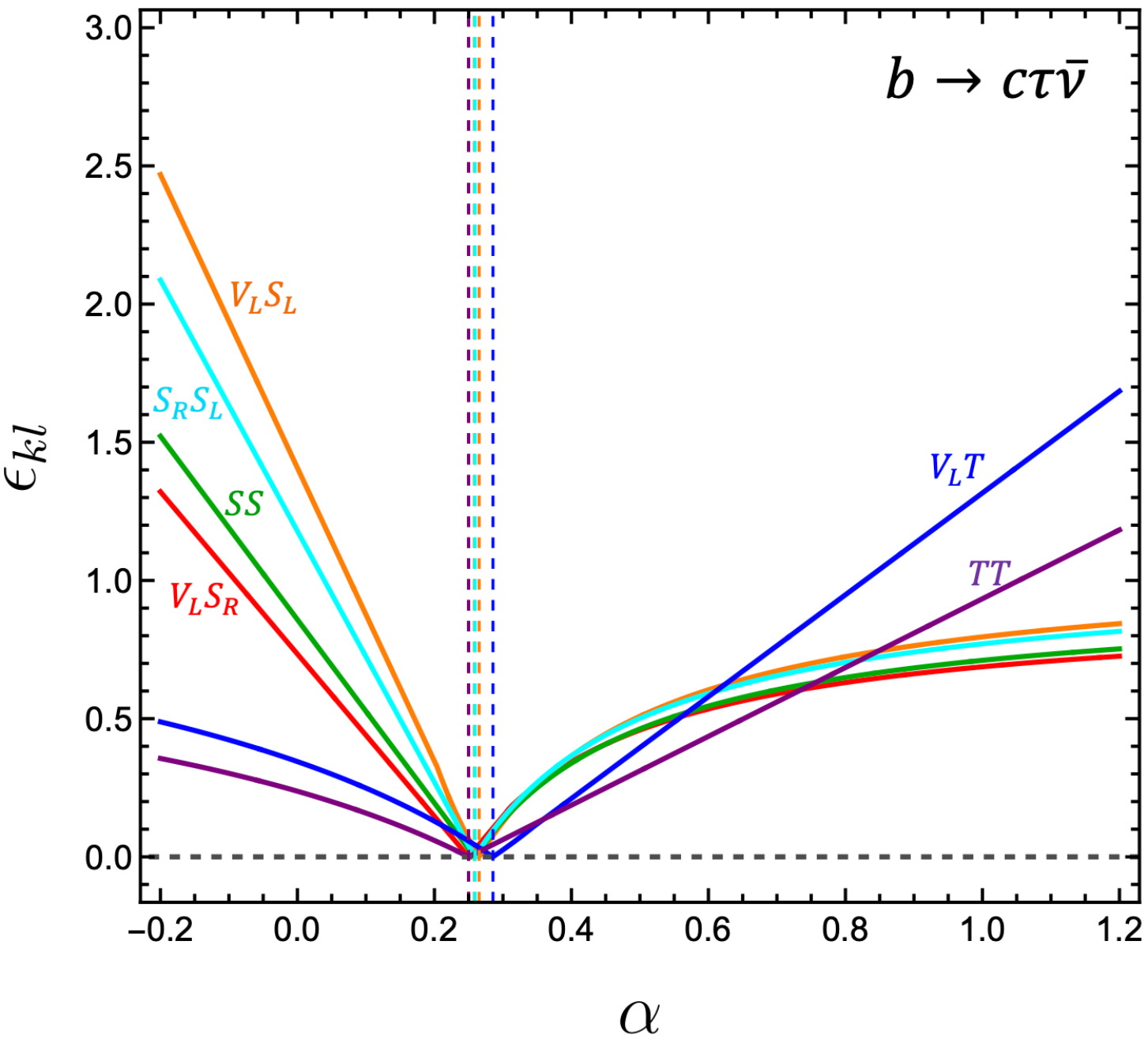}\\\vspace{5mm}
\includegraphics[width=0.448\textwidth]{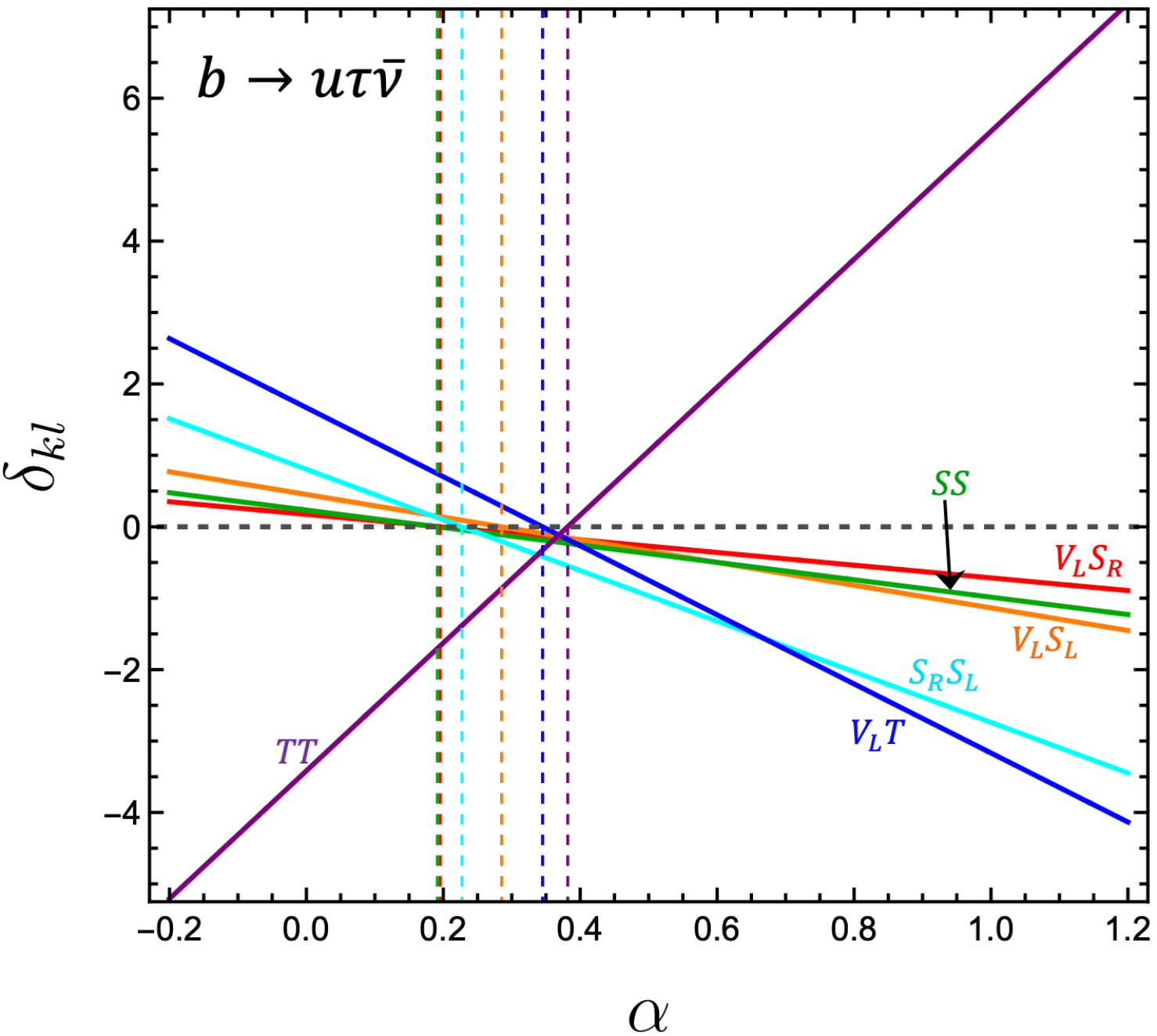}~~~
\includegraphics[width=0.45\textwidth]{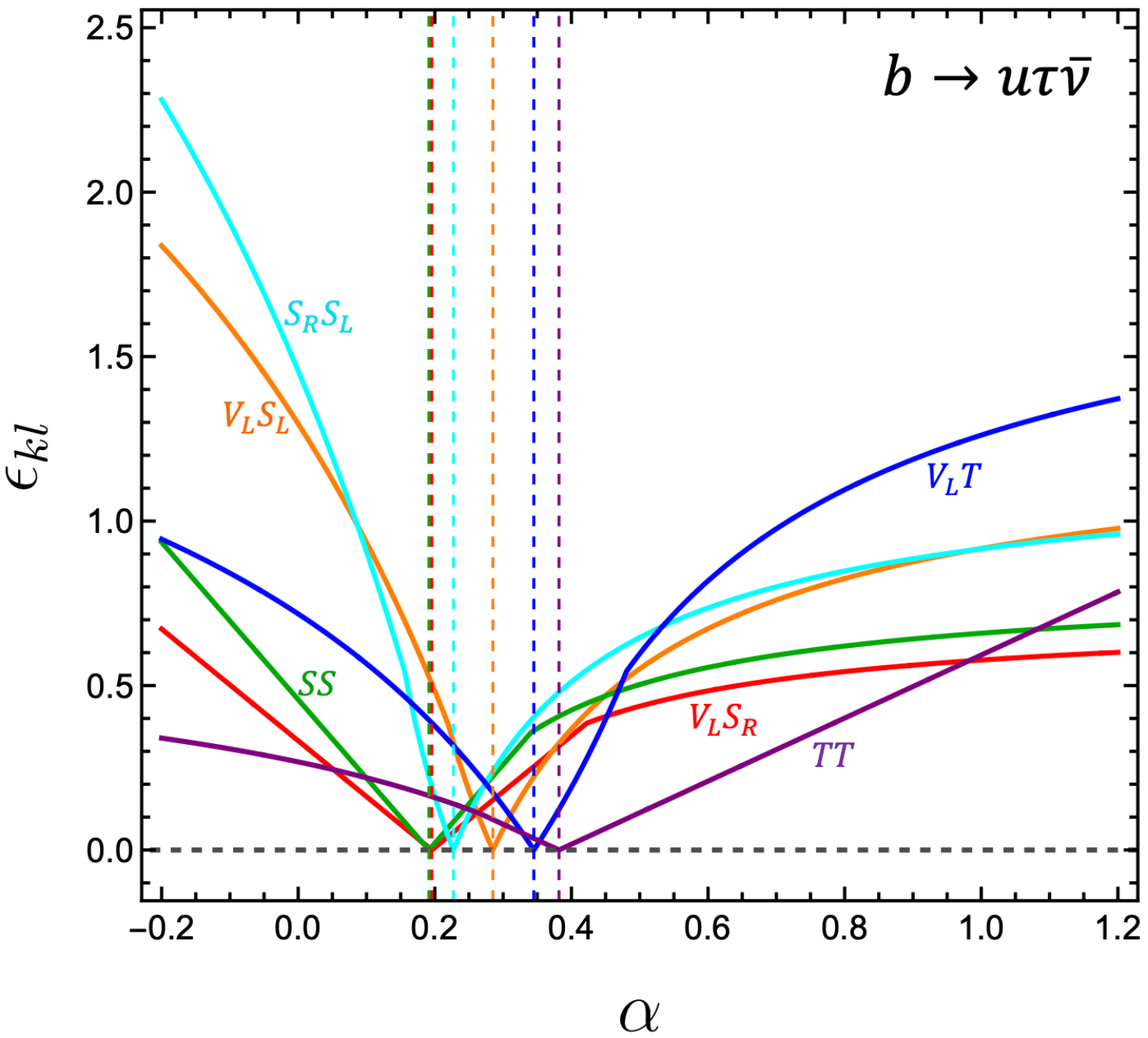}
\caption{
   \label{fig:deleps_alpha_tau}
   Same as Fig.~\ref{fig:deleps_alpha}, but for $b\to c\tau\ov\nu$ (upper) and $b\to u\tau\ov\nu$ (lower).
   } 
\end{center}
\vspace{-3mm}
\end{figure*}
%%%%%%%%%%%%%%%%%%%%%%%%%%%%%%%%%%%%%%%%%%%%%%%%%

Figure~\ref{fig:deleps_alpha_tau} shows the corresponding results for the tauonic bottom-hadron decays.
The values of $\alpha_{mn}$ are more closely aligned than in the corresponding muonic modes.
For $b\to c$ transitions, this behavior can be understood from the small-velocity limit, where the sum-rule coefficient approaches $\alpha = 1/4$ \cite{Endo:2025lvy}.
Deviations from this value arise from higher-order corrections in the heavy-quark expansion and from departures from zero recoil.
The latter effect is more pronounced for light leptons, since the kinematically allowed recoil range is larger than that in the tauonic mode.
Equivalently, with $w=(m_{H_b}^2+m_{H_c}^2-q^2)/(2m_{H_b}m_{H_c})$, one has $w_{\rm max}^\tau<w_{\rm max}^\mu$.
A related discussion is given in Ref.~\cite{Endo:2026dxr}.

%%%%%%%%%%%%%%%%%%%%%%%%%%%%%%%%%%%%%
\bibliographystyle{utphys28mod}
\bibliography{ref}

@article{Kitahara:2026doj,
    author = "Kitahara, Teppei and Mohapatra, Manas Kumar and Sasaki, Kota",
    title = "{Baryon-Meson Sum Rule for $b \to s \nu\bar \nu$}",
    eprint = "2604.20340",
    archivePrefix = "arXiv",
    primaryClass = "hep-ph",
    reportNumber = "CHIBA-EP-278",
    month = "4",
    year = "2026"
}

@article{Lee:2025kvf,
    author = "Lee, Jong-Phil",
    title = "{$\Lambda_b\to \Lambda^{(*)}\nu{\bar \nu}$ and $b\to s$$B$ decays}",
    eprint = "2509.26370",
    archivePrefix = "arXiv",
    primaryClass = "hep-ph",
    month = "9",
    year = "2025"
}

@article{Riggio:2017zwh,
    author = "Riggio, L. and Salerno, G. and Simula, S.",
    title = "{Extraction of $|V_{cd}|$ and $|V_{cs}|$ from experimental decay rates using lattice QCD $D \to \pi(K) \ell \nu$ form factors}",
    eprint = "1706.03657",
    archivePrefix = "arXiv",
    primaryClass = "hep-lat",
    reportNumber = "PREPRINT-RM3-TH-17-7, preprint RM3-TH/17-7",
    doi = "10.1140/epjc/s10052-018-5943-5",
    journal = "Eur. Phys. J. C",
    volume = "78",
    number = "6",
    pages = "501",
    year = "2018"
}

@article{Leljak:2023gna,
    author = "Leljak, Domagoj and Meli{\'c}, Bla{\v{z}}enka and Novak, Filip and Reboud, M{\'e}ril and van Dyk, Danny",
    title = "{Toward a complete description of $b\to u\ell^{-} \overline{\nu}$ decays within the Weak Effective Theory}",
    eprint = "2302.05268",
    archivePrefix = "arXiv",
    primaryClass = "hep-ph",
    reportNumber = "EOS-2023-01, IPPP/23/01, TUM-HEP-1452/23, RBI-ThPhys-2023-3",
    doi = "10.1007/JHEP08(2023)063",
    journal = "JHEP",
    volume = "08",
    pages = "063",
    year = "2023"
}

@article{Leskovec:2025gsw,
    author = "Leskovec, Luka and Meinel, Stefan and Petschlies, Marcus and Negele, John and Paul, Srijit and Pochinsky, Andrew",
    title = "{$B\to\rho\ell\bar\nu$ Resonance Form Factors from $B\to\pi\pi\ell\bar\nu$ in Lattice QCD}",
    eprint = "2501.00903",
    archivePrefix = "arXiv",
    primaryClass = "hep-lat",
    doi = "10.1103/PhysRevLett.134.161901",
    journal = "Phys. Rev. Lett.",
    volume = "134",
    number = "16",
    pages = "161901",
    year = "2025"
}

@inproceedings{Iguro:2026xgi,
    author = "Iguro, Syuhei",
    title = "{$b \to c$ semileptonic sum rule: SU(3)$_{\rm{F}}$ symmetry violation}",
    booktitle = "{60th Rencontres de Moriond on QCD and High Energy Interactions}: {Moriond QCD 2026}",
    eprint = "2604.05186",
    archivePrefix = "arXiv",
    primaryClass = "hep-ph",
    month = "4",
    year = "2026"
}

@article{Iguro:2026,
    author = "Endo, Motoi and Iguro, Syuhei and Mishima, Satoshi",
    title = "{$b\to c$ semileptonic sum rule: orbitally excited hadrons}",
    eprint = "2604.27970",
    archivePrefix = "arXiv",
    primaryClass = "hep-ph",
    reportNumber = "KEK-TH-2828",
    month = "4",
    year = "2026"
}

@article{Endo:2025lvy,
    author = "Endo, Motoi and Iguro, Syuhei and Mishima, Satoshi and Watanabe, Ryoutaro",
    title = "{$b\to c$ semileptonic sum rule: current status and prospects}",
    eprint = "2508.06322",
    archivePrefix = "arXiv",
    primaryClass = "hep-ph",
    reportNumber = "KEK-TH-2744",
    doi = "10.1007/JHEP01(2026)143",
    journal = "JHEP",
    volume = "01",
    pages = "143",
    year = "2026"
}

@article{Endo:2026dxr,
    author = "Endo, Motoi and Iguro, Syuhei and Kretz, Tim and Mishima, Satoshi",
    title = "{b {\textrightarrow} c semileptonic sum rule: exploring a sterile neutrino loophole}",
    eprint = "2603.15029",
    archivePrefix = "arXiv",
    primaryClass = "hep-ph",
    reportNumber = "KEK-TH-2818, P3H-26-016, TTP26-005",
    doi = "10.1007/JHEP06(2026)260",
    journal = "JHEP",
    volume = "06",
    pages = "260",
    year = "2026"
}

@article{Endo:2025fke,
    author = "Endo, Motoi and Iguro, Syuhei and Mishima, Satoshi and Watanabe, Ryoutaro",
    title = "{Heavy quark symmetry behind $b\to c$ semileptonic sum rule}",
    eprint = "2501.09382",
    archivePrefix = "arXiv",
    primaryClass = "hep-ph",
    reportNumber = "KEK-TH-2680, KEK--TH--2680",
    doi = "10.1007/JHEP05(2025)112",
    journal = "JHEP",
    volume = "05",
    pages = "112",
    year = "2025"
}

@article{Endo:2025cvu,
    author = "Endo, Motoi and Iguro, Syuhei and Kretz, Tim and Mishima, Satoshi and Watanabe, Ryoutaro",
    title = "{$b \rightarrow c$ semileptonic sum rule: extension to angular observables}",
    eprint = "2506.16027",
    archivePrefix = "arXiv",
    primaryClass = "hep-ph",
    reportNumber = "KEK-TH-2729, P3H-25-038, TTP25-018",
    doi = "10.1140/epjc/s10052-025-14598-9",
    journal = "Eur. Phys. J. C",
    volume = "85",
    number = "9",
    pages = "961",
    year = "2025",
    note = "[Erratum: Eur.Phys.J.C 85, 1050 (2025)]"
}

@article{Endo:2025set,
    author = "Endo, Motoi and Iguro, Syuhei and Mishima, Satoshi and Watanabe, Ryoutaro",
    title = "{Constructing heavy-quark sum rule for $b\to c$ meson and baryon decays}",
    eprint = "2509.02006",
    archivePrefix = "arXiv",
    primaryClass = "hep-ph",
    reportNumber = "KEK-TH-2754",
    doi = "10.1103/xs7x-jcrd",
    journal = "Phys. Rev. D",
    volume = "113",
    number = "1",
    pages = "014021",
    year = "2026"
}

@article{BESIII:2018nzb,
    author = "Ab-likim, M. and others",
    collaboration = "BESIII",
    title = "{Measurement of the branching fraction for the semi-leptonic decay $D^{0(+)}\to \pi^{-(0)}\mu^+\nu_\mu$ and test of lepton universality}",
    eprint = "1802.05492",
    archivePrefix = "arXiv",
    primaryClass = "hep-ex",
    doi = "10.1103/PhysRevLett.121.171803",
    journal = "Phys. Rev. Lett.",
    volume = "121",
    number = "17",
    pages = "171803",
    year = "2018"
}

@article{CMS:2018hff,
    author = "Sirunyan, Albert M and others",
    collaboration = "CMS",
    title = "{Search for high-mass resonances in final states with a lepton and missing transverse momentum at $ \sqrt{s}=13 $ TeV}",
    eprint = "1803.11133",
    archivePrefix = "arXiv",
    primaryClass = "hep-ex",
    reportNumber = "CMS-EXO-16-033, CERN-EP-2018-020",
    doi = "10.1007/JHEP06(2018)128",
    journal = "JHEP",
    volume = "06",
    pages = "128",
    year = "2018"
}

@article{Abitalk,
    author = "Soffer, Abner",
    collaboration = "Belle II",
    title = "{Recent updates and future prospects of the Belle II experiment}",
    month = "2",
    year = "2025",
    note = "{\url{https://conference-indico.kek.jp/event/278/contributions/6384/}}"
}

@article{Crivellin:2025qsq,
    author = "Crivellin, Andreas and Iguro, Syuhei and Kitahara, Teppei",
    title = "{Discriminating tauphilic leptoquark explanations of the B anomalies via $K\to\pi\nu\bar\nu$ and $B\to K\nu\bar\nu$}",
    eprint = "2505.05552",
    archivePrefix = "arXiv",
    primaryClass = "hep-ph",
    reportNumber = "ZU-TH 32/25, KEK-TH-2718, CHIBA-EP-271",
    doi = "10.1103/4dpx-h5vm",
    journal = "Phys. Rev. D",
    volume = "112",
    number = "9",
    pages = "095016",
    year = "2025"
}

@article{Belle-II:2025ruy,
    author = "Adachi, I. and others",
    collaboration = "Belle-II",
    title = "{Measurement of $B^+\to \tau^+\nu_\tau$ branching fraction with a hadronic tagging method at Belle II}",
    eprint = "2502.04885",
    archivePrefix = "arXiv",
    primaryClass = "hep-ex",
    reportNumber = "Belle II preprint 2024-033, KEK preprint 2024-48",
    doi = "10.1103/dcwd-5tg4",
    journal = "Phys. Rev. D",
    volume = "112",
    number = "7",
    pages = "072002",
    year = "2025"
}

@article{Belle:2015qal,
    author = "Hamer, P. and others",
    collaboration = "Belle",
    title = "{Search for $B^0 \to \pi^- \tau^+ \nu_\tau$ with hadronic tagging at Belle}",
    eprint = "1509.06521",
    archivePrefix = "arXiv",
    primaryClass = "hep-ex",
    reportNumber = "KEK-PREPRINT-2015-23",
    doi = "10.1103/PhysRevD.93.032007",
    journal = "Phys. Rev. D",
    volume = "93",
    number = "3",
    pages = "032007",
    year = "2016"
}

@article{ATLAS:2019lsy,
    author = "Aad, Georges and others",
    collaboration = "ATLAS",
    title = "{Search for a heavy charged boson in events with a charged lepton and missing transverse momentum from $pp$ collisions at $\sqrt{s} = 13$ TeV with the ATLAS detector}",
    eprint = "1906.05609",
    archivePrefix = "arXiv",
    primaryClass = "hep-ex",
    reportNumber = "CERN-EP-2019-100",
    doi = "10.1103/PhysRevD.100.052013",
    journal = "Phys. Rev. D",
    volume = "100",
    number = "5",
    pages = "052013",
    year = "2019"
}

@article{Glashow:1961tr,
    author = "Glashow, S. L.",
    title = "{Partial Symmetries of Weak Interactions}",
    doi = "10.1016/0029-5582(61)90469-2",
    journal = "Nucl. Phys.",
    volume = "22",
    pages = "579--588",
    year = "1961"
}

@article{Weinberg:1967tq,
    author = "Weinberg, Steven",
    title = "{A Model of Leptons}",
    doi = "10.1103/PhysRevLett.19.1264",
    journal = "Phys. Rev. Lett.",
    volume = "19",
    pages = "1264--1266",
    year = "1967"
}

@article{Salam:1968rm,
    author = "Salam, Abdus",
    title = "{Weak and Electromagnetic Interactions}",
    doi = "10.1142/9789812795915_0034",
    journal = "Conf. Proc. C",
    volume = "680519",
    pages = "367--377",
    year = "1968"
}

@article{FCC:2025lpp,
    author = "Benedikt, M. and others",
    collaboration = "FCC",
    title = "{Future Circular Collider Feasibility Study Report: Volume 1, Physics, Experiments, Detectors}",
    eprint = "2505.00272",
    archivePrefix = "arXiv",
    primaryClass = "hep-ex",
    reportNumber = "CERN-FCC-PHYS-2025-0002",
    doi = "10.1140/epjc/s10052-025-15077-x",
    journal = "Eur. Phys. J. C",
    volume = "85",
    number = "12",
    pages = "1468",
    year = "2025"
}

@article{Achasov:2023gey,
    author = "Achasov, M. and others",
    title = "{STCF conceptual design report (Volume 1): Physics {\&} detector}",
    eprint = "2303.15790",
    archivePrefix = "arXiv",
    primaryClass = "hep-ex",
    doi = "10.1007/s11467-023-1333-z",
    journal = "Front. Phys. (Beijing)",
    volume = "19",
    number = "1",
    pages = "14701",
    year = "2024"
}

@article{Cerri:2018ypt,
    author = "Cerri, A. and others",
    editor = "Dainese, Andrea and Mangano, Michelangelo and Meyer, Andreas B. and Nisati, Aleandro and Salam, Gavin and Vesterinen, Mika Anton",
    title = "{Report from Working Group 4}: {Opportunities in Flavour Physics at the HL-LHC and HE-LHC}",
    eprint = "1812.07638",
    archivePrefix = "arXiv",
    primaryClass = "hep-ph",
    reportNumber = "CERN-LPCC-2018-06",
    doi = "10.23731/CYRM-2019-007.867",
    journal = "CERN Yellow Rep. Monogr.",
    volume = "7",
    pages = "867--1158",
    year = "2019"
}

@article{BESIII:2020nme,
    author = "Ablikim, M. and others",
    collaboration = "BESIII",
    title = "{Future Physics Programme of BESIII}",
    eprint = "1912.05983",
    archivePrefix = "arXiv",
    primaryClass = "hep-ex",
    reportNumber = "HEP-Physics-Report-BESIII-2019-12-13",
    doi = "10.1088/1674-1137/44/4/040001",
    journal = "Chin. Phys. C",
    volume = "44",
    number = "4",
    pages = "040001",
    year = "2020"
}

@article{Fuentes-Martin:2020lea,
    author = "Fuentes-Martin, Javier and Greljo, Admir and Martin Camalich, Jorge and Ruiz-Alvarez, Jos{\'e} David",
    title = "{Charm physics confronts high-$p_{T}$ lepton tails}",
    eprint = "2003.12421",
    archivePrefix = "arXiv",
    primaryClass = "hep-ph",
    reportNumber = "CERN-TH-2020-047, ZU-TH 07/20",
    doi = "10.1007/JHEP11(2020)080",
    journal = "JHEP",
    volume = "11",
    pages = "080",
    year = "2020"
}

@article{Meinel:2017ggx,
    author = "Meinel, Stefan",
    title = "{$\Lambda_c \to N$ form factors from lattice QCD and phenomenology of $\Lambda_c \to n \ell^+ \nu_\ell$ and $\Lambda_c \to p \mu^+ \mu^-$ decays}",
    eprint = "1712.05783",
    archivePrefix = "arXiv",
    primaryClass = "hep-lat",
    reportNumber = "RBRC-1262, RBRC-1262",
    doi = "10.1103/PhysRevD.97.034511",
    journal = "Phys. Rev. D",
    volume = "97",
    number = "3",
    pages = "034511",
    year = "2018"
}

@article{HeavyFlavorAveragingGroupHFLAV:2024ctg,
    author = "Banerjee, Sw. and others",
    collaboration = "Heavy Flavor Averaging Group (HFLAV)",
    title = "{Averages of b-hadron, c-hadron, and {\ensuremath{\tau}}-lepton properties as of 2023}",
    eprint = "2411.18639",
    archivePrefix = "arXiv",
    primaryClass = "hep-ex",
    doi = "10.1103/x87q-tld5",
    journal = "Phys. Rev. D",
    volume = "113",
    number = "1",
    pages = "012008",
    year = "2026"
}

@article{Sakaki:2013bfa,
    author = "Sakaki, Yasuhito and Tanaka, Minoru and Tayduganov, Andrey and Watanabe, Ryoutaro",
    title = "{Testing leptoquark models in $\bar B \to D^{(*)} \tau \bar\nu$}",
    eprint = "1309.0301",
    archivePrefix = "arXiv",
    primaryClass = "hep-ph",
    reportNumber = "OU-HET-791, KEK-TH-1660, OU-HET 791",
    doi = "10.1103/PhysRevD.88.094012",
    journal = "Phys. Rev. D",
    volume = "88",
    number = "9",
    pages = "094012",
    year = "2013"
}

@article{Datta:2017aue,
    author = "Datta, Alakabha and Kamali, Saeed and Meinel, Stefan and Rashed, Ahmed",
    title = "{Phenomenology of $ {\Lambda}_b\to {\Lambda}_c\tau {\overline{\nu}}_{\tau } $ using lattice QCD calculations}",
    eprint = "1702.02243",
    archivePrefix = "arXiv",
    primaryClass = "hep-ph",
    doi = "10.1007/JHEP08(2017)131",
    journal = "JHEP",
    volume = "08",
    pages = "131",
    year = "2017"
}

@article{Duan:2024ayo,
    author = "Duan, Wen-Feng and Iguro, Syuhei and Li, Xin-Qiang and Watanabe, Ryoutaro and Yang, Ya-Dong",
    title = "{On sum rules for semi-leptonic $b\to c$ and $b \to u$ decays}",
    eprint = "2410.21384",
    archivePrefix = "arXiv",
    primaryClass = "hep-ph",
    doi = "10.1007/JHEP07(2025)166",
    journal = "JHEP",
    volume = "07",
    pages = "166",
    year = "2025"
}

@article{Blanke:2019qrx,
    author = "Blanke, Monika and Crivellin, Andreas and Kitahara, Teppei and Moscati, Marta and Nierste, Ulrich and Ni\v{s}and\v{z}i\'c, Ivan",
    title = "{Addendum to \textquotedblleft{}Impact of polarization observables and $B_c\to \tau \nu$ on new physics explanations of the $b\to c \tau \nu$ anomaly''}",
    eprint = "1905.08253",
    archivePrefix = "arXiv",
    primaryClass = "hep-ph",
    reportNumber = "PSI-PR-19-09; ZU-TH 26/19; TTP-19-012; P3H-19-011",
    doi = "10.1103/PhysRevD.100.035035",
    month = "5",
    year = "2019",
    note = "[Addendum: Phys.Rev.D 100, 035035 (2019)]"
}

@article{Blanke:2018yud,
    author = "Blanke, Monika and Crivellin, Andreas and de Boer, Stefan and Kitahara, Teppei and Moscati, Marta and Nierste, Ulrich and Ni\v{s}and\v{z}i\'c, Ivan",
    title = "{Impact of polarization observables and $ B_c\to \tau \nu$ on new physics explanations of the $b\to c \tau \nu$ anomaly}",
    eprint = "1811.09603",
    archivePrefix = "arXiv",
    primaryClass = "hep-ph",
    reportNumber = "PSI-PR-18-16; TTP-18-42, PSI-PR--18--16, TTP--18--42",
    doi = "10.1103/PhysRevD.99.075006",
    journal = "Phys. Rev. D",
    volume = "99",
    number = "7",
    pages = "075006",
    year = "2019"
}

@article{Iguro:2020keo,
    author = "Iguro, Syuhei and Takeuchi, Michihisa and Watanabe, Ryoutaro",
    title = "{Testing leptoquark/EFT in ${\bar{B}} \rightarrow {D^{(*)}}l{\bar{\nu }}$ at the LHC}",
    eprint = "2011.02486",
    archivePrefix = "arXiv",
    primaryClass = "hep-ph",
    doi = "10.1140/epjc/s10052-021-09125-5",
    journal = "Eur. Phys. J. C",
    volume = "81",
    number = "5",
    pages = "406",
    year = "2021"
}

@article{Fedele:2022iib,
    author = "Fedele, Marco and Blanke, Monika and Crivellin, Andreas and Iguro, Syuhei and Kitahara, Teppei and Nierste, Ulrich and Watanabe, Ryoutaro",
    title = "{Impact of $\Lambda_b\to\Lambda_c \tau\nu$ measurement on new physics in $b\to c l\nu$ transitions}",
    eprint = "2211.14172",
    archivePrefix = "arXiv",
    primaryClass = "hep-ph",
    reportNumber = "PSI-PR-22-34, ZU-TH 56/22, TTP22-069, P3H-22-113, KEK-TH-2474",
    doi = "10.1103/PhysRevD.107.055005",
    journal = "Phys. Rev. D",
    volume = "107",
    number = "5",
    pages = "055005",
    year = "2023"
}

@article{PDG2024,
    author = "Navas, S. and others",
    collaboration = "Particle Data Group",
    title = "{Review of particle physics}",
    doi = "10.1103/PhysRevD.110.030001",
    journal = "Phys. Rev. D",
    volume = "110",
    number = "3",
    pages = "030001",
    year = "2024"
}

@article{Bernlochner:2018bfn,
    author = "Bernlochner, Florian U. and Ligeti, Zoltan and Robinson, Dean J. and Sutcliffe, William L.",
    title = "{Precise predictions for $\Lambda_b \to \Lambda_c$ semileptonic decays}",
    eprint = "1812.07593",
    archivePrefix = "arXiv",
    primaryClass = "hep-ph",
    doi = "10.1103/PhysRevD.99.055008",
    journal = "Phys. Rev. D",
    volume = "99",
    number = "5",
    pages = "055008",
    year = "2019"
}

@article{Belle-II:2018jsg,
    author = "Altmannshofer, W. and others",
    editor = "Kou, E. and Urquijo, P.",
    collaboration = "Belle-II",
    title = "{The Belle II Physics Book}",
    eprint = "1808.10567",
    archivePrefix = "arXiv",
    primaryClass = "hep-ex",
    reportNumber = "KEK Preprint 2018-27, BELLE2-PUB-PH-2018-001, FERMILAB-PUB-18-398-T, JLAB-THY-18-2780, INT-PUB-18-047, UWThPh 2018-26",
    doi = "10.1093/ptep/ptz106",
    journal = "PTEP",
    volume = "2019",
    number = "12",
    pages = "123C01",
    year = "2019",
    note = "[Erratum: PTEP 2020, 029201 (2020)]"
}

@article{Iguro:2020cpg,
    author = "Iguro, Syuhei and Watanabe, Ryoutaro",
    title = "{Bayesian fit analysis to full distribution data of $\bar B \to D^{(*)} \ell\bar\nu$: $|V_{cb}|$ determination and New Physics constraints}",
    eprint = "2004.10208",
    archivePrefix = "arXiv",
    primaryClass = "hep-ph",
    doi = "10.1007/JHEP08(2020)006",
    journal = "JHEP",
    volume = "08",
    pages = "006",
    year = "2020"
}

@article{Lubicz:2017syv,
    author = "Lubicz, V. and Riggio, L. and Salerno, G. and Simula, S. and Tarantino, C.",
    collaboration = "ETM",
    title = "{Scalar and vector form factors of $D \to \pi(K) \ell \nu$ decays with $N_f=2+1+1$ twisted fermions}",
    eprint = "1706.03017",
    archivePrefix = "arXiv",
    primaryClass = "hep-lat",
    reportNumber = "PREPRINT-RM3-TH-17-6, preprint RM3-TH/17-6",
    doi = "10.1103/PhysRevD.96.054514",
    journal = "Phys. Rev. D",
    volume = "96",
    number = "5",
    pages = "054514",
    year = "2017",
    note = "[Erratum: Phys.Rev.D 99, 099902 (2019), Erratum: Phys.Rev.D 100, 079901 (2019)]"
}

@article{Lubicz:2018rfs,
    author = "Lubicz, V. and Riggio, L. and Salerno, G. and Simula, S. and Tarantino, C.",
    collaboration = "ETM",
    title = "{Tensor form factor of $D \to \pi(K) \ell \nu$ and $D \to \pi(K) \ell \ell$ decays with $N_f=2+1+1$ twisted-mass fermions}",
    eprint = "1803.04807",
    archivePrefix = "arXiv",
    primaryClass = "hep-lat",
    doi = "10.1103/PhysRevD.98.014516",
    journal = "Phys. Rev. D",
    volume = "98",
    number = "1",
    pages = "014516",
    year = "2018"
}

@article{Wu:2006rd,
    author = "Wu, Yue-Liang and Zhong, Ming and Zuo, Ya-Bing",
    title = "{$B_s$, $D_s$ $\to$ $\pi$, $K$, $\eta$, $\rho$, $K^*$, $\omega$, $\phi$ Transition Form Factors and Decay Rates with Extraction of the CKM parameters $|V_{ub}|$, $|V_{cs}|$, $|V_{cd}|$}",
    eprint = "hep-ph/0604007",
    archivePrefix = "arXiv",
    doi = "10.1142/S0217751X06033209",
    journal = "Int. J. Mod. Phys. A",
    volume = "21",
    pages = "6125--6172",
    year = "2006"
}

@article{BESIII:2024lxg,
    author = "Ablikim, Medina and others",
    collaboration = "BESIII",
    title = "{Study of the decay $D^0\to\rho(770)^-e^+\nu_e$}",
    eprint = "2409.04276",
    archivePrefix = "arXiv",
    primaryClass = "hep-ex",
    doi = "10.1103/PhysRevD.110.112018",
    journal = "Phys. Rev. D",
    volume = "110",
    number = "11",
    pages = "112018",
    year = "2024"
}

@article{BESIII:2018qmf,
    author = "Ablikim, Medina and others",
    collaboration = "BESIII",
    title = "{Observation of $D^+ \to f_0(500) e^+\nu_e$ and Improved Measurements of $D \to\rho e^+\nu_e$}",
    eprint = "1809.06496",
    archivePrefix = "arXiv",
    primaryClass = "hep-ex",
    doi = "10.1103/PhysRevLett.122.062001",
    journal = "Phys. Rev. Lett.",
    volume = "122",
    number = "6",
    pages = "062001",
    year = "2019"
}

@article{CLEO:2011ab,
    author = "Dobbs, S. and others",
    collaboration = "CLEO",
    title = "{First Measurement of the Form Factors in the Decays $D^0 \to \rho^- e^+ \nu_e$ and $D^+ \to \rho^0 e^+ \nu_e$}",
    eprint = "1112.2884",
    archivePrefix = "arXiv",
    primaryClass = "hep-ex",
    reportNumber = "CLNS-11-2075, CLEO-11-3",
    doi = "10.1103/PhysRevLett.110.131802",
    journal = "Phys. Rev. Lett.",
    volume = "110",
    number = "13",
    pages = "131802",
    year = "2013"
}

@article{CLEO:2005cuk,
    author = "Coan, T. E. and others",
    collaboration = "CLEO",
    title = "{Absolute branching fraction measurements of exclusive $D^0$ semileptonic decays}",
    eprint = "hep-ex/0506052",
    archivePrefix = "arXiv",
    reportNumber = "CLNS-05-1906, CLEO-05-01",
    doi = "10.1103/PhysRevLett.95.181802",
    journal = "Phys. Rev. Lett.",
    volume = "95",
    pages = "181802",
    year = "2005"
}

@article{BESIII:2021pvy,
    author = "Ablikim, Medina and others",
    collaboration = "BESIII",
    title = "{Observation of the decay $D^0\to \rho^-\mu^+\nu_\mu$}",
    eprint = "2106.02292",
    archivePrefix = "arXiv",
    primaryClass = "hep-ex",
    reportNumber = "BAM-00467",
    doi = "10.1103/PhysRevD.104.L091103",
    journal = "Phys. Rev. D",
    volume = "104",
    number = "9",
    pages = "L091103",
    year = "2021"
}

@article{BESIII:2015tql,
    author = "Ablikim, M. and others",
    collaboration = "BESIII",
    title = "{Study of Dynamics of $D^0 \to K^- e^+ \nu_{e}$ and $D^0\to\pi^- e^+ \nu_{e}$ Decays}",
    eprint = "1508.07560",
    archivePrefix = "arXiv",
    primaryClass = "hep-ex",
    doi = "10.1103/PhysRevD.92.072012",
    journal = "Phys. Rev. D",
    volume = "92",
    number = "7",
    pages = "072012",
    year = "2015"
}

@article{CLEO:2009svp,
    author = "Besson, D. and others",
    collaboration = "CLEO",
    title = "{Improved measurements of $D$ meson semileptonic decays to $\pi$ and $K$ mesons}",
    eprint = "0906.2983",
    archivePrefix = "arXiv",
    primaryClass = "hep-ex",
    reportNumber = "CLNS-09-2049, CLEO-09-02",
    doi = "10.1103/PhysRevD.80.032005",
    journal = "Phys. Rev. D",
    volume = "80",
    pages = "032005",
    year = "2009"
}

@article{CLEO:2007ntr,
    author = "Dobbs, S. and others",
    collaboration = "CLEO",
    title = "{A Study of the semileptonic charm decays $D^0 \to \pi^- e^+ \nu_e$, $D^+ \to \pi^0 e^+ \nu_e$, $D^0 \to K^- e^+ \nu_e$, and $D^+ \to \bar{K}^0 e^+ \nu_e$}",
    eprint = "0712.1020",
    archivePrefix = "arXiv",
    primaryClass = "hep-ex",
    reportNumber = "CLNS-06-1968, CLEO-06-13",
    doi = "10.1103/PhysRevD.77.112005",
    journal = "Phys. Rev. D",
    volume = "77",
    pages = "112005",
    year = "2008"
}

@article{Belle:2006idb,
    author = "Widhalm, L. and others",
    collaboration = "Belle",
    title = "{Measurement of $D^0 \to \pi l \nu (Kl \nu)$ Form Factors and Absolute Branching Fractions}",
    eprint = "hep-ex/0604049",
    archivePrefix = "arXiv",
    reportNumber = "BELLE-PREPRINT-2006-12",
    doi = "10.1103/PhysRevLett.97.061804",
    journal = "Phys. Rev. Lett.",
    volume = "97",
    pages = "061804",
    year = "2006"
}

@article{Bowler:1994zr,
    author = "Bowler, K. C. and Hazel, N. M. and Hoeber, H. and Kenway, R. D. and Richards, D. G. and Lellouch, L. and Nieves, J. and Sachrajda, Christopher T. and Wittig, H.",
    collaboration = "UKQCD",
    title = "{An 'improved' lattice study of semileptonic decays of D mesons}",
    eprint = "hep-lat/9410012",
    archivePrefix = "arXiv",
    reportNumber = "EDINBURGH-94-546, SHEP-93-94-32",
    doi = "10.1103/PhysRevD.51.4905",
    journal = "Phys. Rev. D",
    volume = "51",
    pages = "4905--4923",
    year = "1995"
}

@article{FlavourLatticeAveragingGroupFLAG:2024oxs,
    author = "Aoki, Y. and others",
    collaboration = "Flavour Lattice Averaging Group (FLAG)",
    title = "{FLAG review 2024}",
    eprint = "2411.04268",
    archivePrefix = "arXiv",
    primaryClass = "hep-lat",
    reportNumber = "CERN-TH-2024-192, FERMILAB-PUB-24-0785-T",
    doi = "10.1103/nfzp-p5dn",
    journal = "Phys. Rev. D",
    volume = "113",
    number = "1",
    pages = "014508",
    year = "2026"
}

@article{Bazavov:2017lyh,
    author = "Bazavov, A. and others",
    title = "{$B$- and $D$-meson leptonic decay constants from four-flavor lattice QCD}",
    eprint = "1712.09262",
    archivePrefix = "arXiv",
    primaryClass = "hep-lat",
    reportNumber = "FERMILAB-PUB-17/491-T, FERMILAB-PUB-17-491-T",
    doi = "10.1103/PhysRevD.98.074512",
    journal = "Phys. Rev. D",
    volume = "98",
    number = "7",
    pages = "074512",
    year = "2018"
}

@article{Carrasco:2014poa,
    author = "Carrasco, N. and others",
    title = "{Leptonic decay constants $f_{K},f_{D},$ and $f_{{D}_{s}}$ with $N_{f} = 2+1+1$ twisted-mass lattice QCD}",
    eprint = "1411.7908",
    archivePrefix = "arXiv",
    primaryClass = "hep-lat",
    doi = "10.1103/PhysRevD.91.054507",
    journal = "Phys. Rev. D",
    volume = "91",
    number = "5",
    pages = "054507",
    year = "2015"
}

@article{Fu:2018yin,
    author = {Fu, Hai-Bing and Zeng, Long and L{\"u}, Rong and Cheng, Wei and Wu, Xing-Gang},
    title = "{The $D\to \rho$ semileptonic and radiative decays within the light-cone sum rules}",
    eprint = "1808.06412",
    archivePrefix = "arXiv",
    primaryClass = "hep-ph",
    doi = "10.1140/epjc/s10052-020-7758-4",
    journal = "Eur. Phys. J. C",
    volume = "80",
    number = "3",
    pages = "194",
    year = "2020"
}

@article{Lin:2025cmn,
    author = "Lin, Wang and Huang, Xiao-En and Cheng, Shan and Yao, De-Liang",
    title = "{Semileptonic decays of $D\to\rho l^+\nu$ and $D_{(s)}\to K^*l^+\nu$ from light-cone sum rules}",
    eprint = "2505.01329",
    archivePrefix = "arXiv",
    primaryClass = "hep-ph",
    doi = "10.1103/jp8y-j9g8",
    journal = "Phys. Rev. D",
    volume = "111",
    number = "11",
    pages = "113005",
    year = "2025"
}
\end{document}